\documentclass[10pt,conference]{IEEEtran}
\IEEEoverridecommandlockouts
\usepackage{cite}
\usepackage{amsmath,amssymb,amsfonts}
\usepackage{algorithmic}
\usepackage{graphicx}
\usepackage{textcomp}
\usepackage[table,xcdraw]{xcolor}
\usepackage{fancyhdr}
\usepackage{soul}
\soulregister{\cite}{7}
\soulregister{\cref}{7}
\soulregister{\ref}{7}
\soulregister{\eqref}{7}
\soulregister{\section}{7}
\soulregister{\subsection}{7}
\soulregister{\subsubsection}{7}
\soulregister{\sysname}{7}
\soulregister{\textsc}{7}
\soulregister{\textbf}{7}
\soulregister{\xspace}{7}
\soulregister{\footnote}{7}
\soulregister{\item}{7}
\soulregister{\underline}{7}
\soulregister{\ie}{7}

\renewcommand{\hl}[1]{#1}

\usepackage{makecell}
\usepackage{booktabs}
\usepackage{xspace}
\usepackage[shortlabels]{enumitem}
\usepackage[ruled,linesnumbered]{algorithm2e}
\usepackage{subcaption}
\usepackage{multirow}
\usepackage{graphicx}
\usepackage{multicol}
\usepackage{caption}
\usepackage{subcaption}
\usepackage{xcolor}
\definecolor{orange}{HTML}{BE6548}
\definecolor{purple}{HTML}{72346E}
\usepackage[bookmarks=true,breaklinks=true,letterpaper=true,colorlinks,citecolor=blue,linkcolor=blue,urlcolor=blue]{hyperref}

\usepackage{tikz}
\newcommand{\ballnumber}[1]{\tikz[baseline=(myanchor.base)] \node[circle,fill=.,inner sep=1pt] (myanchor) {\color{white}{\bfseries\footnotesize #1}};}

\makeatletter
\DeclareRobustCommand\onedot{\futurelet\@let@token\@onedot}
\def\@onedot{\ifx\@let@token.\else.\null\fi\xspace}

\def\eg{\emph{e.g}\onedot} 
\def\ie{\emph{i.e}\onedot} 
 
 \def\vs{\emph{vs}\onedot}
\def\wrt{w.r.t\onedot} 

\def\aka{\emph{aka}\onedot} 
\makeatother

\newlength\savewidth

\usepackage{pifont}
\newcommand{\cmark}{\ding{51}}%
\newcommand{\xmark}{\ding{55}}%

\usepackage{amsmath,amsfonts,bm}

\def\eqref#1{equation~\ref{#1}}

\def\1{\bm{1}}

\def\va{{\bm{a}}}

\def\vc{{\bm{c}}}
\def\vd{{\bm{d}}}

\def\vs{{\bm{s}}}

\def\vv{{\bm{v}}}
\def\vw{{\bm{w}}}

\DeclareMathAlphabet{\mathsfit}{\encodingdefault}{\sfdefault}{m}{sl}
\SetMathAlphabet{\mathsfit}{bold}{\encodingdefault}{\sfdefault}{bx}{n}

\usepackage[capitalise]{cleveref}

\crefformat{section}{#2Sec.~#1#3}
\crefformat{figure}{#2Fig.~#1#3}
\crefformat{table}{#2Tab.~#1#3}
\crefformat{theorem}{#2Theorem~#1#3}
\crefformat{lemma}{#2Lemma~#1#3}
\crefname{section}{sec.}{secs.}
\crefname{figure}{fig.}{figs.}
\crefname{table}{tab.}{tabs.}
\crefname{theorem}{thm.}{thms.}
\crefname{lemma}{lem.}{lems.}

\usepackage{booktabs}
\usepackage{tabularx}
\usepackage{array}
\usepackage{multirow}

\newcommand{\mytablesize}{8}
\newcommand{\mytablebaselineskip}{9}
\newcolumntype{Y}{>{\centering\arraybackslash}X}

\usepackage{xspace}
\newcommand{\sysname}{\textsc{ELSA}\xspace}

\def\BibTeX{{\rm B\kern-.05em{\sc i\kern-.025em b}\kern-.08em
    T\kern-.1667em\lower.7ex\hbox{E}\kern-.125emX}}
\begin{document}

\pdfpagewidth=8.5in
\pdfpageheight=11in

\newcommand{\iscasubmissionnumber}{124}

\pagenumbering{arabic}


\title{\sysname: An \underline{EL}astic \underline{S}NN Inference \underline{A}rchitecture for Efficient Neuromorphic Computing
\thanks{
The work is supported in part by the National Key R\&D Program of China under Grant 2022YFB4500200 and Shanghai Artificial Intelligence Laboratory.
}
}

\makeatletter
\newcommand{\linebreakand}{%
  \end{@IEEEauthorhalign}
  \hfill\mbox{}\par
  \mbox{}\hfill\begin{@IEEEauthorhalign}
}
\makeatother

\author{
\IEEEauthorblockN{
Kang You\IEEEauthorrefmark{1}\IEEEauthorrefmark{2},
Chen Nie\IEEEauthorrefmark{1}\IEEEauthorrefmark{2},
Lee Jun Yan\IEEEauthorrefmark{1},
Ziling Wei\IEEEauthorrefmark{1}\IEEEauthorrefmark{2},
Cheng Zou\IEEEauthorrefmark{1},
Zekai Xu\IEEEauthorrefmark{1},\\
Yu Feng\IEEEauthorrefmark{3},
Honglan Jiang\IEEEauthorrefmark{4},
Zhezhi He\IEEEauthorrefmark{1}\IEEEauthorrefmark{2}\IEEEauthorrefmark{5}
}
\IEEEauthorblockA{\IEEEauthorrefmark{1}Intelligent Computing Research Group, School of Computer Science, Shanghai Jiao Tong University, Shanghai, CN
}
\IEEEauthorblockA{
\IEEEauthorrefmark{2}Shanghai AI Laboratory, Shanghai, CN,
\IEEEauthorrefmark{3}School of Computer Science, Shanghai Jiao Tong University, Shanghai, CN
}
\IEEEauthorblockA{\IEEEauthorrefmark{4}Institute of Chip Design and EDA, School of Integrated Circuits, Shanghai Jiao Tong University, Shanghai, CN
}
\IEEEauthorblockA{\IEEEauthorrefmark{5}Corresponding author
\vspace{-2em}
}
}

\maketitle
\thispagestyle{plain}
\pagestyle{plain}

\begin{abstract}

Spiking neural networks (SNNs) exploit event-driven and addition-only computation to substantially improve efficiency for intelligent computation.
A key temporal property of SNNs, elastic inference, allows outputs to emerge progressively, enabling responses to salient inputs much earlier than full evaluation.
However, existing SNN-specific accelerators cannot capitalize on this property.
Layer-by-layer designs emit outputs only after all layers are complete, while time-step-by-time-step designs rely on coarse-grained, layer-wise pipelines that require synchronizing all spines/tokens within a layer.
This barrier prevents results from being forwarded immediately, delaying the earliest possible response and forfeiting the benefits of elastic inference.

To address these challenges, we propose \sysname, \textit{a near-SRAM dataflow architecture that realizes true elastic inference through a fine-grained spine/token-wise pipeline and hardware optimizations tailored to SNNs.}
\sysname forwards each spine/token immediately upon production, forming a continuous streaming pipeline that substantially reduces the latency to the first response.
To enhance this lightweight execution, \sysname introduces a bundled address event representation protocol to lower communication traffic of network-on-chip (NoC), and leverages mini-batch spiking Gustavson-product to cut memory access and exploit inherent sparsity.
Combined with mapping and scheduling optimizations, \sysname achieves efficient, event-driven computation without compromising accuracy.
Experiments show that \textit{SNNs can outperform quantized artificial neural networks (QANNs) while maintaining on-par accuracy.} 
For a 4-bit ResNet-50, \sysname achieves 3.4$\times$ speedup and 13.6$\times$ higher energy efficiency over the SOTA QANN accelerator (ANT), and 2.9$\times$ speedup and 22.1$\times$ energy efficiency gains over the SOTA SNN accelerator (PAICORE).
\end{abstract}

\section{Introduction}
\label{sec:introduction}
Spiking neural networks (SNNs) \cite{bu2023optimal, hu2023fast, zhou2022spikformer, spikeziptf2024} encode information using discrete binary or ternary spikes, closely mimicking the dynamics of biological neurons.
Compared with artificial neural networks (ANNs), SNNs feature event-driven and addition-only computation \cite{roy2019towards} with notably higher activation sparsity (\eg, 98\% \cite{fang2024energy}), enabling efficient processing \cite{C-DNN, liu2022sato, yin2024loas, lee2022parallel, akopyan2015truenorth, davies2018loihi, PAICORE, darwin3}.
This work focuses on a previously unexploited property of SNNs, \emph{elastic inference}, unlocking new opportunities to further boost the performance of SNN-specific hardware.

\begin{figure}[t]
    \centering
    \begin{subfigure}{\linewidth}
        \includegraphics[width=\linewidth]{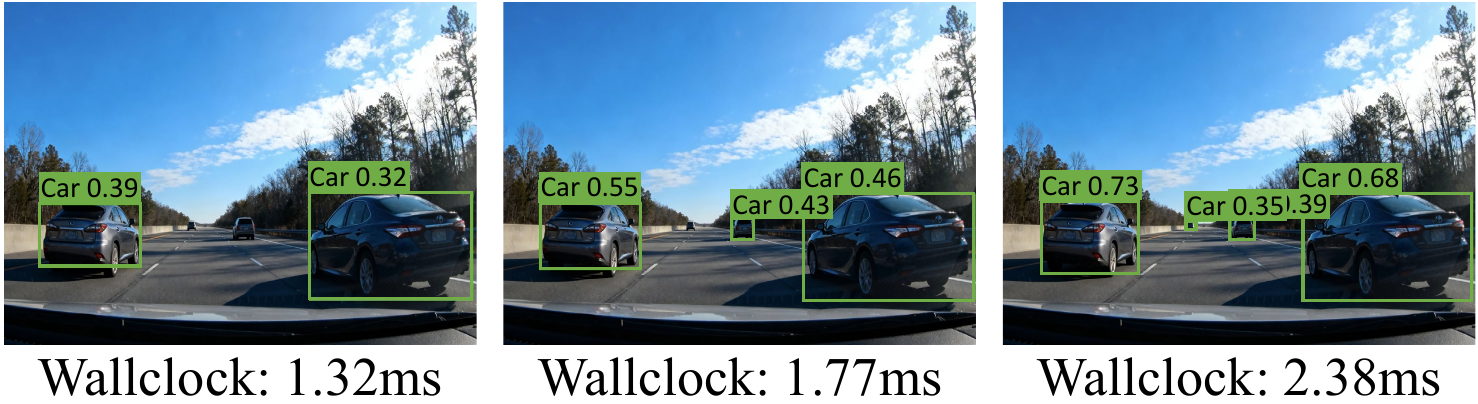}
        \caption{Elastic inference enables early detection of salient objects.}
        \label{fig:subfig1a}
    \end{subfigure}
    \\
    \begin{subfigure}{\linewidth}
    \includegraphics[width=\linewidth]{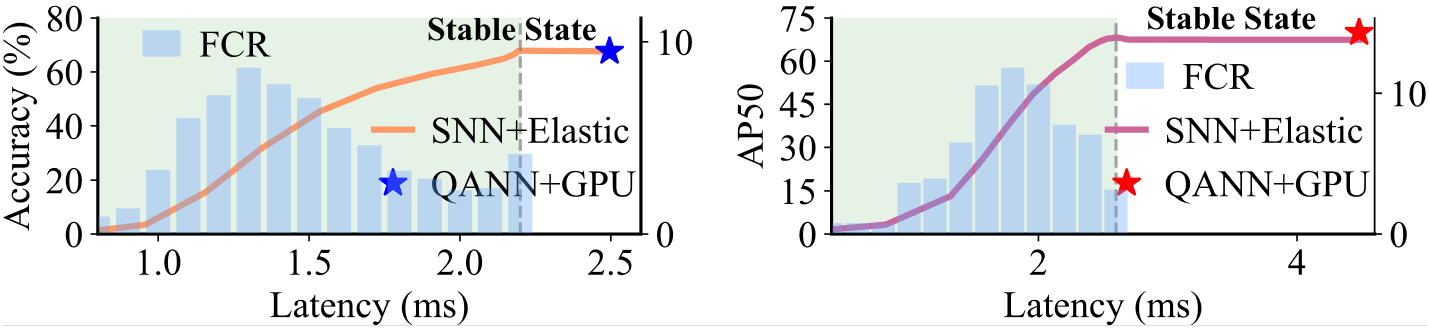}
        \caption{
        Classification accuracy on ImageNet and detection AP50 on COCO2017 improve progressively with latency.}
        \label{fig:subfig1b}
    \end{subfigure}
    \caption{
    \textbf{Illustration of elastic inference.} 
    Bars denote first-correct-response (FCR) latency, dashed lines mark stable-state outputs, and stars show QANN execution on an A100 GPU.
    }
    \vspace{-1em}
    \label{fig:characteristic_of_SNN}
\end{figure}

\emph{Elastic inference} is a unique temporal property of SNNs, where outputs emerge progressively, allowing earlier responses to salient inputs.
Given sufficient inference time, the final predictions converge to those obtained from full execution.
For instance, in \cref{fig:subfig1a}, visually prominent vehicles are recognized earlier, while distant ones require additional inference time.
This phenomenon is consistent with early decision-making in biological neural systems \cite{thorpe1996speed}, where \emph{salient stimuli trigger faster neural responses}.
Such temporal elasticity is particularly valuable for real-time tasks such as autonomous driving \cite{qian20223d}, where the first correct response can arrive up to 82\% earlier than the stable-state output, as shown in \cref{fig:subfig1b}.

\begin{figure*}[t]
\centering
    \includegraphics[width=\linewidth]{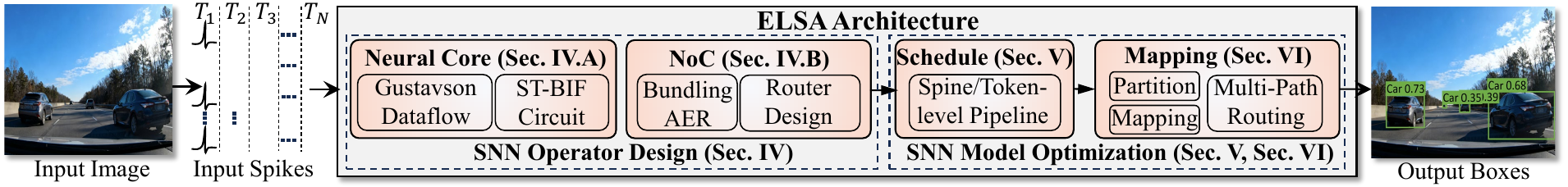}
    \vspace{-1.5em}
    \caption{
    \textbf{Overall architecture and execution flow of \sysname}.
    }
    \label{fig:ELSA_organization}
    \vspace{-1.5em}
\end{figure*}

Nevertheless, existing SNN accelerators \cite{liu2022sato, yin2024loas, lee2022parallel, akopyan2015truenorth, davies2018loihi, morphic, PAICORE, darwin3} barely exploit elastic inference.
Their execution can be broadly categorized as \textit{layer-by-layer} (LBL) and \textit{time-step-by-time-step} (TBT), distinguished by how they traverse the three intrinsic dimensions of SNNs: time-steps\footnote{A time-step~\cite{spikeziptf2024} is a discrete interval in which synaptic transmissions occur and neurons integrate inputs and generate spikes once.}, layers, and spines/tokens within each layer (defined in \cref{fig:token_spine_in_SNN}).
LBL-based accelerators \cite{liu2022sato, yin2024loas, lee2022parallel} process all time-steps of one layer before moving to the next, producing outputs only after the full network completes. Thus, they are inherently incompatible with elastic inference.
TBT-based accelerators \cite{akopyan2015truenorth,davies2018loihi,morphic,PAICORE,darwin3} 
evaluate all layers at every time-step, thus allowing progressively emerging outputs and supporting elastic inference.

However, existing TBT-based accelerators \cite{akopyan2015truenorth, davies2018loihi, morphic, PAICORE, darwin3} still adopt coarse-grained layer-wise pipelines, where computation advances only after all $N$ spines/tokens in a layer are buffered and synchronized.
This prevents completed spines/tokens from being forwarded immediately. Consequently, early responses can emerge only after the final layer is reached.
Such layer-level synchronization substantially delays the first possible response compared with an ideal fine-grained spine/token-wise pipeline, as illustrated in \cref{fig:pipeline_comparison}.
These limitations motivate an SNN accelerator that supports true fine-grained spine/token-wise pipelining for elastic inference.

\begin{figure}[t]
\includegraphics[width=\columnwidth]{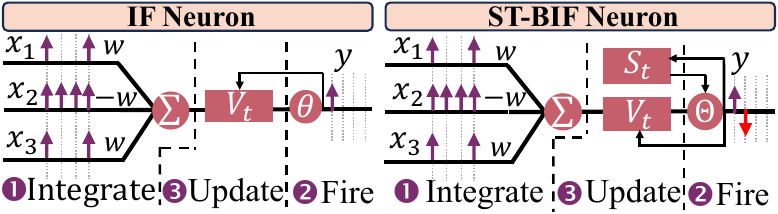}
\caption{Neural dynamics of (left) IF and (right) ST-BIF neuron.}
\label{fig:neural_dynamic}
\vspace{-1.5em}
\end{figure}

This work proposes \sysname, a near-SRAM dataflow architecture that exploits elastic inference in convolutional and transformer-based SNNs.
\sysname introduces a fine-grained spine/token-wise pipeline, supported by dedicated mapping algorithms, that allows each completed spine/token to advance to the next layer immediately.
This fine-grained processing is essential for achieving low-latency elastic inference.
\cref{fig:ELSA_organization} illustrates the overall architecture and execution flow.
Input spikes encoded across time-steps are continuously processed by neural
cores and delivered through the network-on-chip (NoC).
To further improve efficiency, \sysname incorporates SNN-specific hardware
optimizations, including bundled AER for compact spike communication and
mini-batch spiking Gustavson-product for sparse, memory-efficient computation.
The key contributions are as follows:
\begin{itemize}[leftmargin=*,label={$\triangleright$}] 

\item \textbf{Fine-grained pipeline for elastic inference:}
\sysname introduces a spine/token-wise pipeline that forwards each completed
spine/token to the next layer immediately.
Such streaming execution exploits elastic inference to reduce the latency to the first response and improve throughput.

\item \textbf{SNN-aware hardware optimizations:}
We propose a \textit{bundled AER} protocol that packs multiple spikes into one flit, reducing NoC traffic under fine-grained pipelining.
We further design a \textit{mini-batch spiking Gustavson-product} dataflow to reduce memory accesses while exploiting spike sparsity.
Together, these optimizations lower on-chip communication and computation energy.

\item \textbf{High-performance elastic inference:} 
\sysname is the first SNN accelerator explicitly optimized for elastic inference.
It achieves on-par accuracy while delivering a 3.4$\times$ speedup and 13.6$\times$ energy savings over the SOTA QANN accelerator (ANT \cite{ANT}), and 2.9$\times$ speedup and 22.1$\times$ energy savings over the SOTA SNN accelerator (PAICORE\cite{PAICORE}).

\end{itemize}

\section{Preliminary of SNN}
\label{sec:snn_basic}

\subsection{Neural Dynamics of Spiking Neurons}

\subsubsection{Integrate-and-Fire (IF) Neuron} 
\label{sec:spiking_neuron}
The IF neuron is widely used in SNNs and has been adopted
by many neuromorphic chips \cite{akopyan2015truenorth, davies2018loihi, C-DNN, mao2024stellar}. 
Unlike continuous activations (\eg, ReLU), IF neurons communicate via binary spike trains (\{0,1\}), enabling event-driven and addition-only computation.
However, IF-based SNNs incur an accuracy loss relative to ANN
counterparts due to conversion errors \cite{hu2023fast, spikeziptf2024}.

\subsubsection{Bipolar Integrate-and-Fire with Spike Tracer (ST-BIF) Neuron}
\label{sec:background_STBIF}
The ST-BIF neuron can be mathematically equivalent to quantized ReLU under specific conditions \cite{spikeziptf2024}.
As shown in \cref{fig:neural_dynamic}, ST-BIF emits ternary spikes
(\{-1,0,1\}) via three steps:

\textbf{Step-1: Spikes Integration.}
The neuron receives and integrates pre-synaptic spikes $x_{i,t} \in \{-1,0,1\}$ into the membrane $V_t$ at $t$ time-step through synaptic weight $w_i$:
\begin{equation}
\begin{array}{c}
\hat{V}_{t} = V_{t-1} + \sum_{i=1}^N x_{i,t} \cdot w_i
\end{array}
\label{eqt:spike-integration}
\end{equation}
where $V_{t-1}$ is the membrane potential before integration, $\hat{V}_{t}$ is the membrane potential after integration, $t$ is the time-step. 

\textbf{Step-2: Neuron Firing.}
After integration, the neuron emits a spike according to the decision function $\Theta$:
\begin{equation}
\label{eqt:ST-BIF-decision}
\begin{gathered}
y_t = \Theta(\hat{V_t}, V_\textrm{thr}, S_{t}) = 
\begin{cases}
1 ;& \hat{V_t} \geq V_\textrm{thr} ~ \& ~ S_t < S_{\textrm{max}} \\
0 ;&  \textrm{other} \\
-1 ;&  \hat{V_t} < 0 ~ \& ~ S_t > S_{\textrm{min}} \\
\end{cases}
\end{gathered}
\end{equation}
where $S_t$ is a memory unit in the ST-BIF neuron (\aka, spike tracer) that records the accumulated sum of emitted spikes. 
$S_{\textrm{max}}$ and $S_{\textrm{min}}$ denote its upper and lower bounds, respectively.
$V_\textrm{thr}$ is the firing threshold.

\textbf{Step-3: Membrane Update.}
After firing, the ST-BIF neuron updates its membrane potential and spike tracer:
\begin{equation}
\label{eqt:IF_membrane_update}
V_t = 
\hat{V}_{t} - y_t \cdot V_\textrm{thr} ;\quad S_t = S_{t-1} + \Theta(\hat{V}_{t}, V_\textrm{thr}, S_{t-1}) 
\end{equation}
The membrane update follows the ``soft reset'' rule \cite{han2020rmp}, while the spike tracer is updated by accumulating the emitted spikes.

\begin{table}[t]
\centering
\caption{SNN operators supported by \sysname.}
\label{tab:back_operators}
{\fontsize{\mytablesize}{\mytablebaselineskip}\selectfont
{
\setlength{\tabcolsep}{4pt}
\begin{tabularx}{\columnwidth}{@{}c c Y@{}}
\textbf{Category} & \textbf{Model} & \textbf{Operators} \\
\midrule
\multirow{2}{*}{Matrix Mult.} 
    & CNN         & MM-sc \\
    & Transformer & MM-sc, MM-ss \\
\midrule
\multirow{2}{*}{Miscellaneous} 
    & CNN         & residual addition, image-to-column \\
    & Transformer & ssoftmax, slayernorm, residual addition \\
\end{tabularx}
}
}
\vspace{-1.0em}
\end{table}

\subsection{Operators in SNN}
\subsubsection{Matrix Multiplication (MM)} 
Unlike conventional MM with two continuous-valued operands, SNNs use spike-continuous MM (MM-sc) and spike-spike MM (MM-ss).
Spiking convolution and linear layers are implemented with MM-sc, while spiking attention uses MM-ss.
Following SpikeZIP-TF \cite{spikeziptf2024}, we realize MM-ss with two MM-sc
operators by treating spike tracers as continuous operands.

\subsubsection{Miscellaneous Operators}
Although MM operators dominate execution time and energy, correct SNN inference
also requires the miscellaneous operators summarized in
\cref{tab:back_operators}.
We follow SpikeZIP-TF \cite{spikeziptf2024} for spiking softmax (ssoftmax) and
spiking layer normalization (slayernorm), and implement image-to-column
transformation and residual addition as router-side broadcasts.

\section{Motivations} 
\label{sec:snn_motivation}

\begin{figure}[t]
\includegraphics[width=\columnwidth]{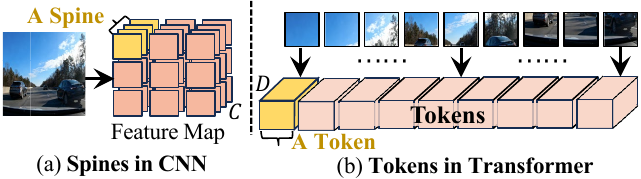}
\caption{\textbf{The definitions of pipeline granularity}. (a) Spine ($\mathbb{Z}^{1\times 1\times C}$) for CNN and (2) Token ($\mathbb{Z}^{1\times D}$) for Transformer.
}
\vspace{-0.5em}
\label{fig:token_spine_in_SNN}
\end{figure}

\begin{figure}[t]
\centering
    \includegraphics[width=\linewidth]{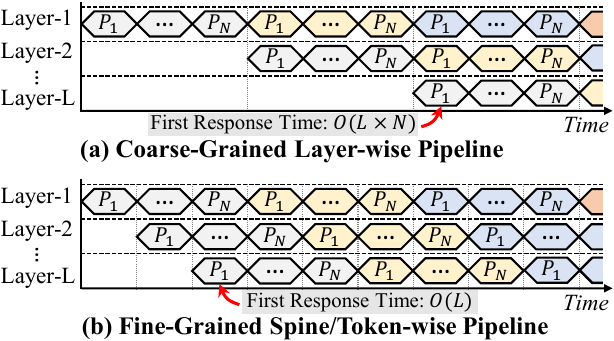}
    \vspace{-1em}
 \caption{\textbf{Comparison of pipeline schemes.} Colors denote different time-steps, and $P_{1\sim N}$ denotes individual spines/tokens.
 The finer-grained pipeline enables substantially earlier first responses, thus better exploiting elastic inference.}
    \label{fig:pipeline_comparison}
    \vspace{-1em}
\end{figure}

\subsection{Fine-Grained Spine/Token-wise Pipeline}
\label{sec:back_pipelining}

\textbf{Problem.}
Elastic inference is an \emph{inherent temporal property of SNNs} that enables correct responses to salient inputs at earlier time-steps.
Unfortunately, existing SNN accelerators \cite{liu2022sato, yin2024loas, lee2022parallel, akopyan2015truenorth, davies2018loihi, C-DNN, PAICORE, darwin3} fail to effectively exploit this property as their execution patterns fundamentally limit temporal elasticity.
Specifically, they follow either a layer-by-layer (LBL) or time-step-by-time-step (TBT) execution pattern, distinguished by how computation traverses the three intrinsic dimensions of SNN inference: time-steps ($T$), layers ($L$), and spines/tokens ($N$) within each layer.
The definitions of spines and tokens are shown in \cref{fig:token_spine_in_SNN}.

LBL-based accelerators \cite{liu2022sato, yin2024loas, lee2022parallel} must complete all $T$ time-steps for all $N$ spines/tokens within a layer (\ie, $T\times N$ computations) before proceeding to the next layer.
As a result, outputs are produced only after the entire network has finished execution, eliminating any opportunity for early responses.
TBT-based SNN accelerators\cite{akopyan2015truenorth, davies2018loihi, PAICORE, darwin3, morphic}, evaluate all $L$ layers at each time-step, enabling progressively generated outputs.
Although these SNN accelerators support elastic inference, their coarse-grained and layer-wise pipeline necessitates the completion of all $N$ spines/tokens before advancing, as shown in \cref{fig:pipeline_comparison}(a).
This design prevents completed spines/tokens from being forwarded immediately, increasing the early response time of elastic inference. 
Overall, existing SNN accelerators fail to effectively exploit elastic inference.

\textbf{Solution.}
We propose a \emph{dataflow architecture with a fine-grained spine/token-wise pipeline that eliminates the coarse synchronization barriers} inherent in prior TBT-based accelerators.
Rather than waiting for all $N$ spines/tokens of a layer to finish, the architecture forwards each completed spine/token to the next layer immediately, regardless of the progress of the others.
As illustrated in
\cref{fig:pipeline_comparison}, this design replaces the conventional coarse-grained layer-wise pipeline (top) with a fine-grained spine/token-wise pipeline (bottom).
The resulting execution forms a continuous streaming pipeline, \textit{reducing the latency to the first response from $O(L\times N)$ to $O(L)$.}
For deep SNNs such as Spikeformer \cite{zhou2022spikformer} with $L = 74$ and $N = 197$, our design can translate into substantial system-level speedup.

\begin{figure}[t]
\centering
    \includegraphics[width=\columnwidth]{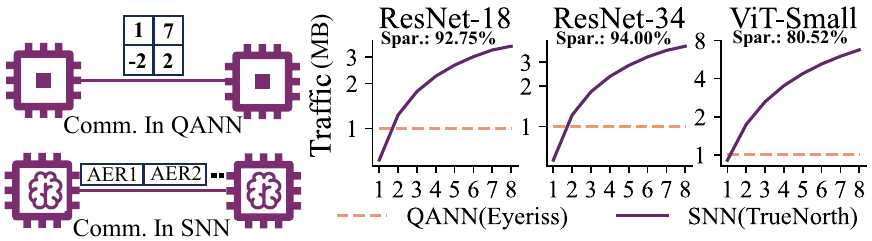}
    \caption{
    Left: Communication comparison of QANN and SNN accelerators.
    Right: TrueNorth traffic relative to Eyeriss \cite{7738524}, using sparsity statistics from SpikeZIP-TF \cite{spikeziptf2024}.
    }
    \label{fig:motivation_traffic}
    \vspace{-0.5em}
\end{figure}

\subsection{Network-on-Chip with Bundled AER Protocol}
\label{sec:network_on_chip}

\textbf{Problem.}
Existing SNN accelerators \cite{davies2018loihi, akopyan2015truenorth, C-DNN, PAICORE, darwin3} exploit the address-event representation (AER) protocol to encode 1-bit spikes as multi-bit packets (\eg, 32 bits in \cite{akopyan2015truenorth}), including the spike's spatial position and time-step information.
Unlike QANN accelerators\cite{7738524} that transmit 8-bit activations, SNN hardware transmits spikes individually over multiple time-steps.
Consequently, even with high spike sparsity (over 80\% in ViTs), TrueNorth \cite{akopyan2015truenorth} can generate up to 8$\times$ more traffic than QANN baselines (\cref{fig:motivation_traffic}).
This overhead arises from large packet headers and repeated transmissions across time-steps, resulting in substantial communication overhead.

\textbf{Solution.}
We introduce a \emph{bundled AER} (BAER) protocol that substantially reduces SNN communication overhead while preserving event-driven behavior.
Instead of transmitting each spike separately, BAER aggregates all spikes produced in the same row of neurons into a single packet, amortizing the header across the group and removing the per-spike header overhead of conventional AER \cite{akopyan2015truenorth}.
This row-wise bundling reduces both packet count and metadata redundancy, yielding a more communication-efficient substrate that aligns naturally with the fine-grained spine/token-wise pipeline of \sysname.

\begin{figure}[t]
\includegraphics[width=\columnwidth]{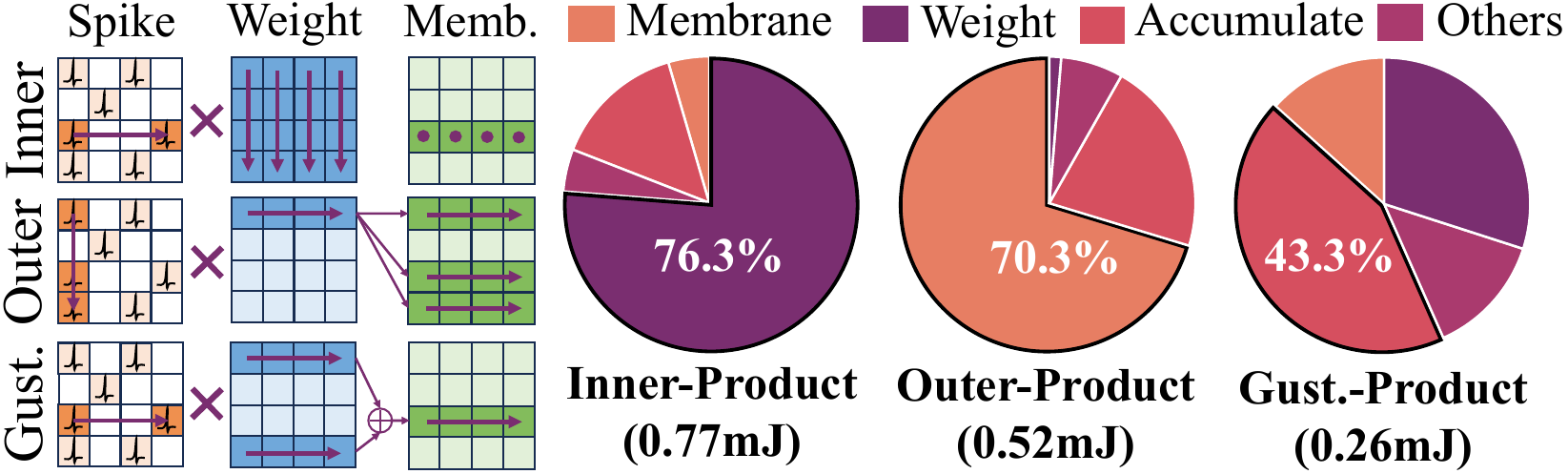}
\caption{\textbf{Energy breakdown when applying different execution patterns to \sysname.} The workload is ResNet-18.}
\label{fig:Gustavson_product}
\vspace{-1.5em}
\end{figure}

\subsection{Neural Core with Mini-Batch Spiking Gustavson-Product}
\label{sec:motication_gustavson}

\textbf{Problem.} 
Gustavson-product \cite{GAMMA} is a widely adopted execution pattern that performs row-wise accumulation, allowing each output row (membrane in SNNs) to be read and written only once, thereby minimizing membrane access.
As illustrated in \cref{fig:Gustavson_product}, Gustavson-product introduces substantially lower memory access than the commonly used inner- and outer-product patterns.
This property is particularly appealing for SNNs, where membranes have a much larger bitwidth (12-bit) than weights (4-bit) or spikes (1-bit).
However, directly applying the ANN-targeted Gustavson-product to SNNs conflicts with the SNN asynchronous feature, where \textit{spikes are forwarded immediately upon generation and thus asynchronously arrive in a row-unaligned order}.
Since Gustavson-product relies on row-aligned inputs to amortize the membrane read/write, such disorder causes frequent switching between membrane rows and largely negates its memory-access benefit.

\textbf{Solution.}
Instead of introducing synchronization \cite{akopyan2015truenorth, davies2018loihi, darwin3} that would break the spine/token-wise pipeline, \sysname adapts Gustavson-product through a mini-batch execution scheme that leverages the row alignment provided by our bundled AER protocol (specified in \cref{sec:network_on_chip}). 
Bundled AER aggregates spikes from the same membrane row into a single flit, forming small row-coherent batches without stalling the pipeline.
Each mini-batch triggers a single read of the corresponding membrane row, parallel accumulation across multiple weight rows, and a single write-back, thereby restoring the low memory access advantage of Gustavson-product.
The resulting mini-batch Gustavson-product produces a steady stream of row-aligned spikes and is therefore naturally compatible with the fine-grained pipeline for elastic inference.

\section{\sysname Architecture}
\label{sec:architecture}

\cref{fig:PE_and_Router}(a) illustrates the scalable architecture of \sysname, which consists of multiple neural cores (dotted frame) interconnected via 2D-mesh NoC \cite{2D_MESH}. 
Each neural core integrates a customized router for communication and four processing elements (PEs) for computation.
Similar to previous TBT-based accelerator \cite{akopyan2015truenorth, davies2018loihi, morphic, PAICORE, darwin3}, \sysname~takes near-SRAM execution, addition-only computation, and event-driven sparsity as the fundamental techniques.

\begin{figure}[t]
\centering
    \includegraphics[width=\linewidth]{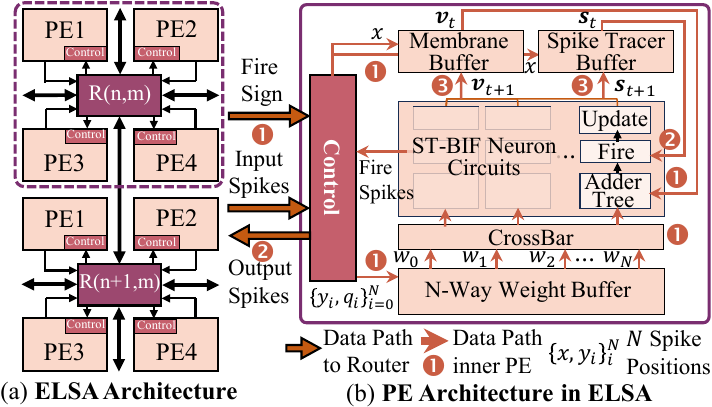}
    \vspace{-1em}
    \caption{\textbf{Overview of \sysname architecture,} consisting of multiple neural cores interconnected by our customized NoC. } 
    \vspace{-.5em}
    \label{fig:PE_and_Router}
\end{figure}

\begin{figure}[t]
    \includegraphics[width=\linewidth]{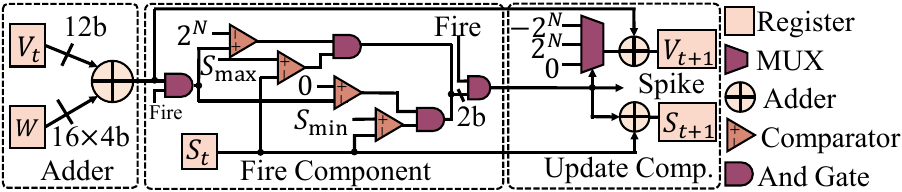}
    \vspace{-1em}
    \caption{\textbf{ST-BIF neuron circuit}, which consists of an adder tree, a fire component, and an update component.} 
    \label{fig:neuron_circuit}
    \vspace{-1.5em}
\end{figure}

\subsection{Microarchitecture of Processing Element}
\label{sec:neural_core}

Our PE is designed to execute MM-sc as listed in \cref{tab:back_operators} via mini-batch spiking Gustavson-product.
As shown in \cref{fig:PE_and_Router}(b), each PE contains 128 ST-BIF neuron circuits, a control module, and massive SRAMs (\ie, an $N$-way weight buffer, a membrane buffer, and a spike tracer buffer).
The ST-BIF neuron circuit, illustrated in \cref{fig:neuron_circuit}, consists of a 16-input adder tree, a fire component, and an update component. In total, each PE can perform 1024 addition operations per cycle.

\begin{figure}[t]
\centering
    \includegraphics[width=\linewidth]{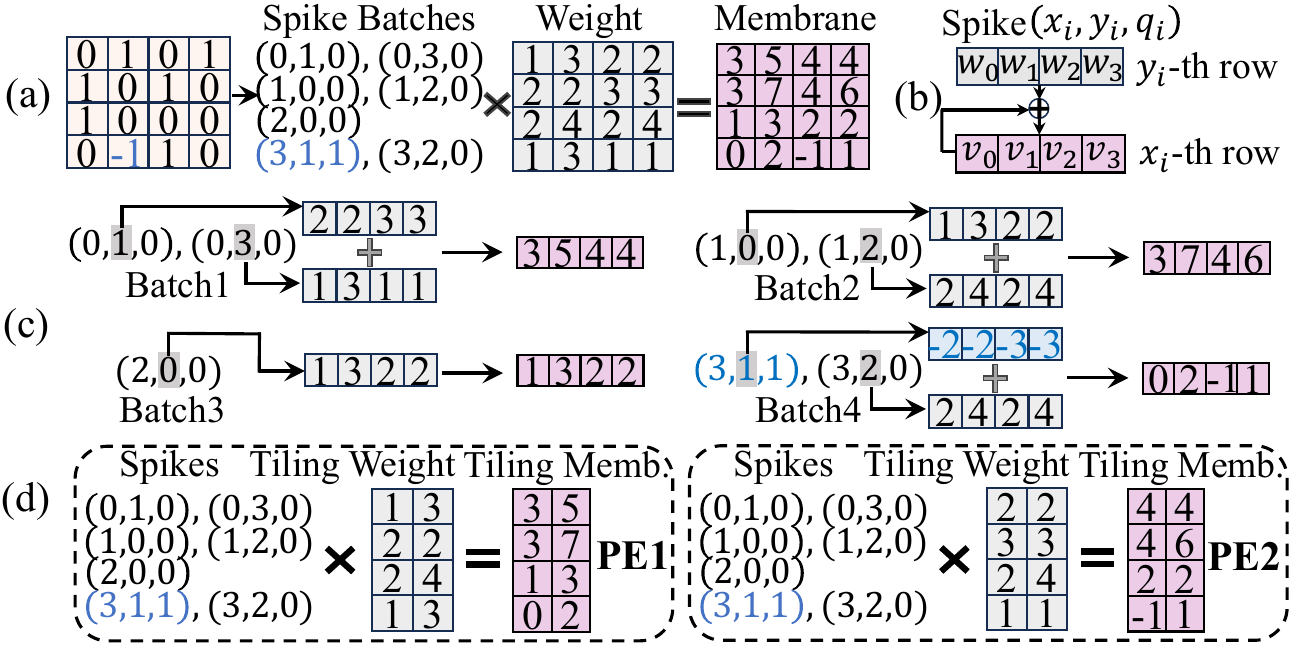}
    \vspace{-1em}
    \caption{\textbf{Process of MM-sc with Mini-batch Gustavson-Product.} (a) The MM-sc in \sysname. (b). Operations per Spike. (c). An example of MM-sc. (d) Tiling Strategy. The process of negative spikes is highlighted in blue.}    
    \label{fig:SMM}
\vspace{-1.5em}
\end{figure}

\subsubsection{Operations per Spike} 
\label{sec:op_per_spike}
For each incoming spike, the PE executes three steps in the ST-BIF neuron (\cref{sec:background_STBIF}), including 1) spike integration, 2) neuron firing, and 3) updating. The dataflows of three steps are marked by \textcolor{orange}{\ballnumber{1}}, \textcolor{orange}{\ballnumber{2}}, \textcolor{orange}{\ballnumber{3}} in \cref{fig:PE_and_Router}(b), respectively. 
For spike integration (step-\textcolor{orange}{\ballnumber{1}}), the control module receives spikes $\{x, y_i, q_i\}^N_{i=0}$ from the router and extracts the encoded positions $x, y_i$ as SRAM addresses. 
Here, $q_i$ denotes the spike polarity (\ie, $q_i = 1$ for negative and $q_i = 0$ for positive). $x,y$ denotes the spike position ($x^{\rm th}$ row and $y^{\rm th}$ column) in spike matrix.
The control module then sends $\{y_i, q_i\}_{i=0}^N$ to the $N$-way weight buffer and $x$ to the membrane buffer, so that the corresponding weights can be accumulated into membrane through the adder tree.
For neuron firing (step-\textcolor{orange}{\ballnumber{2}}), the firing component reads the spike tracer $s_t$ at address $x$ together with the integrated membrane $\hat{V_t}$, and evaluates the decision function in \cref{eqt:ST-BIF-decision}.
For updating (step-\textcolor{orange}{\ballnumber{3}}), the update component updates the membrane state $\vv_{t+1}$ and spike tracer rows $\vs_{t+1}$ based on the firing result.

\subsubsection{Mini-batch Spiking Gustavson-product in PE}
The per-spike execution described above is illustrated in \cref{fig:SMM}(b).
A spike at $(x_i,y_i)$ causes the $y_i$-th row of synaptic weights $\vw$ to be accumulated into the $x_i$-th row of the membrane state $\vv_t$. 
For a negative spike ($q_i = 1$), the corresponding weight row is negated (using two's complement) before accumulation.
For MM-sc, \sysname processes one BAER packet at a time. 
As shown in step-\textcolor{orange}{\ballnumber{1}}, spikes $\{\{x, y_i, q_i\}^N_{i=0}\}$ share a common row address $x$ but have different column IDs $y_i$. 
This row alignment allows the $N$-way weight buffer to process the $N$ spikes in one cycle, which reads $N$ weight rows according to spike addresses $\{y_i\}_{i=0}^N$ and forwards them to the adder trees. 

\cref{fig:SMM}(c) presents an MM-sc example based on the mini-batch Gustavson-product, with negative-spike handling highlighted in blue. 
The first spike batch $(0,1),(0,3)$ triggers the weight buffer to read the 2\textsuperscript{nd} ([2,2,3,3]) and 4\textsuperscript{th} ([1,3,1,1]) rows. 
Then, the adder tree accumulates these weight rows to produce the 1\textsuperscript{st} row of membrane potential ([3,5,4,4]). 
Finally, the fire component receives the integrated results together with the 1\textsuperscript{st} row of spike tracer, performs spike firing, then writes the updated membrane potential and spike tracer back to the membrane and spike tracer buffer, respectively. 

\subsubsection{MM-sc Tiling} 
\label{sec:tiling}
As illustrated in \cref{fig:SMM} (d), we column-wise divide the synaptic weight and membrane (1\textsuperscript{st} and 2\textsuperscript{nd} column to PE1 and 3\textsuperscript{rd} and 4\textsuperscript{th} column to PE2) rather than dividing them block-wise in traditional accelerators. 
With the tiling strategy, spikes are broadcast to all PEs. 
The synaptic weights and membrane potentials are distributed into PEs without overlapping, thus improving the area utilization.

\subsubsection{Multiple MM-sc in Single Neural Core} 
\sysname can map multiple SNN layers with multiple MM-sc into one neural core. 
When a neural core is assigned $P$ MM-sc, it divides the ST-BIF neuron circuits and memories in PE into $P$ groups and allocates them to perform these allocated MM-sc, respectively.

\begin{figure}[t]
\centering
    \includegraphics[width=\linewidth]{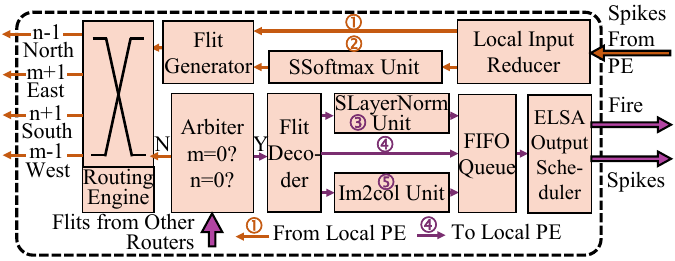}
    \vspace{-1em}
    \caption{\textbf{\sysname Router Design.} \sysname router contains five data paths, two paths \textcolor{orange}{\textcircled{\footnotesize{1}}}  \textcolor{orange}{\textcircled{\footnotesize{2}}} to process spikes from local PEs and three paths \textcolor{purple}{\textcircled{\footnotesize{3}}}  \textcolor{purple}{\textcircled{\footnotesize{4}}} \textcolor{purple}{\textcircled{\footnotesize{5}}} to receive the flits from neural cores. SSoftmax \& SLayerNorm Unit performs the ssoftmax and slayernorm summarized in \cref{tab:back_operators}. $m,n$ are the hop counts in flits (\cref{fig:AER}), $x,y_i$ are the positions in spike matrix.}
    \label{fig:router}
\vspace{-1em}
\end{figure}

\subsection{Router Design and Bundled AER}
\label{sec:router_BAER}
The router in \sysname is responsible for flit generation, communication, and decoding for neighboring neural cores. Note that \textit{our router also supports the execution of miscellaneous multiplication operators} summarized in \cref{tab:back_operators}. 
Moreover, we propose a novel \textit{bundled address-event-representation} (BAER) to reduce the communication traffic compared to vanilla AER.

\subsubsection{Micro-architecture of Router}
\label{sec:router_arch}

As depicted in \cref{fig:router}, the router contains multiple modules and five distinct data paths. Each SNN layer is mapped to one router, with a local path chosen from \textcolor{orange}{\textcircled{\footnotesize{1}}} or \textcolor{orange}{\textcircled{\footnotesize{2}}} for spikes from its PEs and a remote path chosen from \textcolor{purple}{\textcircled{\footnotesize{3}}}, \textcolor{purple}{\textcircled{\footnotesize{4}}}, or \textcolor{purple}{\textcircled{\footnotesize{5}}} for flits from other cores. 
Such an assignment prevents contention across the five data paths.
On the local path, \texttt{Local Input Reducer} gathers spikes until \texttt{Flit Generator} can bundle them into a BAER flit. If there are too few spikes when computation finishes, the flit is zero-padded.
On the remote path, \texttt{Arbiter} monitors each flit’s hop counts $m,n$. When both reach zero, \texttt{Flit Decoder} decodes the flit back to spikes and enqueues them in \texttt{FIFO Queue}. These spikes feed the spine/token-wise pipeline under the control of \texttt{\sysname Output Scheduler}.
For routing, we adopt a static algorithm described in \cref{sec:mapping}, where \texttt{Routing Engine} computes the transmission port for each flit and stores the transmission probability of all ports.

\subsubsection{SNN Operators in Router}
\label{sec:SNN_operators_router}
Router uses \texttt{SSoftmax} and \texttt{SLayerNorm Units} to perform ssoftmax and slayernorm as summarized in \cref{tab:back_operators}. 
We inherit the integer-only softmax and layernorm from \cite{marchisio2023swifttron} to realize ssoftmax and slayernorm. 
Since the outputs of these units are also spikes, a small number of ST-BIF neuron circuits, along with memory units to store neural states, are integrated to support these operators. 
To support convolution layers, \sysname router uses \texttt{im2col Unit} to perform image-to-column\footnote{The image-to-column operator is a transformation used in CNNs, to rearrange image data for efficient matrix multiplications.} broadcasting for each spike.

\begin{figure}[t]
\centering
\includegraphics[width=\linewidth]{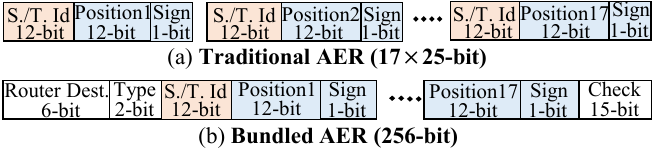}
\vspace{-1em}
\caption{\textbf{(a) Traditional AER and (b) Bundled AER (\aka BAER)}. ``S./T." denotes Spine/Token; ``Dest." is destination. ``Type" is the flit position within a spine/token.}
\label{fig:AER}
\vspace{-1.5em}
\end{figure}

\begin{figure*}[t]
\centering
\includegraphics[width=0.95\linewidth]{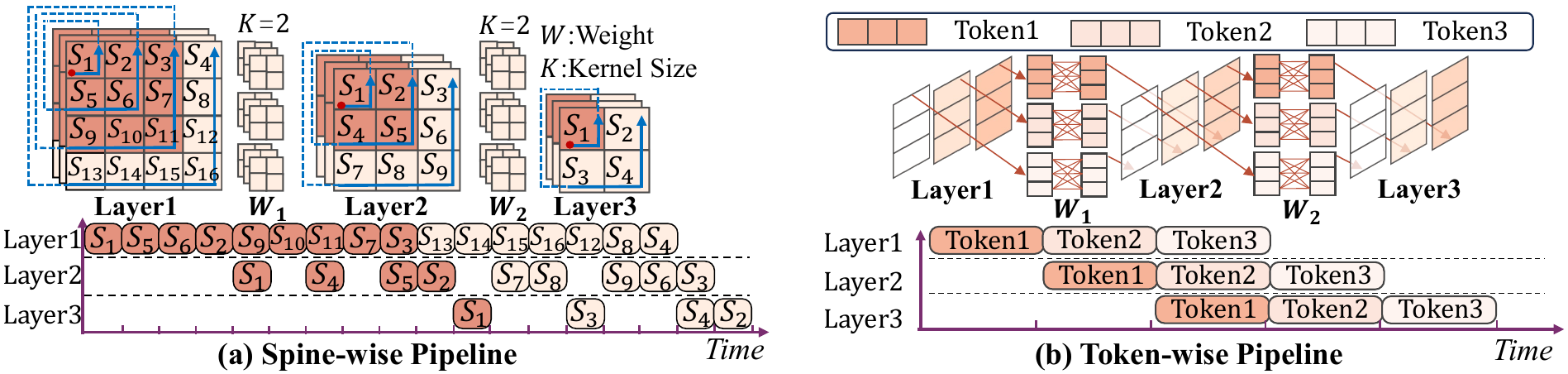}
\caption{
\textbf{Details of fine-grained spine/token-wise pipelines.} 
(a) Spine-wise pipeline in convolution layers. 
The data dependence of the 1\textsuperscript{st} spine ($S_1$) in layer-3 is highlighted in dark orange.
(b) Token-wise pipeline in a multi-layer perceptron.
}
\label{fig:data_dependence}
\vspace{-1em}
\end{figure*}

\subsubsection{Bundled AER (BAER)} 

\cref{fig:AER} highlights the differences when flits are encoded in traditional AER and our BAER. While traditional AER uses independent spine/token ID, position, and sign (\eg, 17 spikes and each consumes 25-bit, thus 425 bits total), BAER can reduce the flit to 256-bit. 
In BAER, the router destination (6-bit) records the hop count ($m$,$n$) for inter-core transmission. The type (2-bit) is the position (\ie, beginning, body, and ending) of the flit within a spine/token. Spine/Token ID (12-bit) is the index for each spine/token. Position (12-bit) is the spike position within a spine/token. Sign (1-bit) is the polarity of the spike. Check (15-bit) records error correcting code for NoC communication.
Last but not least, our BAER naturally aligns with the computation and pipelining granularity in \sysname.

\section{spine/token-wise Pipeline Schedule}
\label{sec:Spine_token_parallelism}
By using mini-batch spiking Gustavson-product (\cref{sec:motication_gustavson}) and bundled AER (\cref{sec:router_BAER}), \sysname explores a spine/token-wise pipeline scheduling to further enhance elastic inference.

\begin{algorithm}[t]
\caption{The control algorithm in \texttt{Output Scheduler} for spine-wise pipeline in CNN.}
\small
\label{alg:control_algorithm}  
Input: kernel height $H_{k}$, kernel width $W_{k}$, convolution stride $S$, convolution padding $P$, height and width of input feature $H_{I}, W_{I}$, position (row, column) of input spine $i,j$.

Output: list $L$ storing the position of output spines.

$i \leftarrow i+P$; $j \leftarrow j+P$; // considering padding

// If all the data-dependent spines arrive.

\If{$i<j$ \textbf{and} $(j-i+1)\!\ge\!H_k$ \textbf{and} $i,j\!\equiv\!0~(\mathrm{mod}~S)$}{
    // The processing order is from right to left.

    $L \leftarrow L \cup \{(i/S,\, (j - W_k + 1)/S)\}$\; 
}

\If{$i>j$ \textbf{and} $i\!\ge\!0$ \textbf{and} $i,j\!\equiv\!0~(\mathrm{mod}~S)$}{
    // The processing order is from bottom to top.

    $L \leftarrow L \cup \{((i - H_k+1)/S,\, (j - W_k + 1)/S)\}$\;
}

// If the last spine arrives, process padding.

\If{$i = P$ \textbf{and} $j = P + W_I - 1$}{
    \For{$p \gets 0$ \textbf{to} $P$}{
        
        $L \leftarrow L \cup \{({ii}/S,\, (j+p - W_k + 1)/S)\}_{ii=0}^{i+p}$\;

        $L \leftarrow L \cup \{({(i+p)}/S,\, (jj - W_k + 1)/S)\}_{jj=0}^{j+p}$\;

    }
}
\end{algorithm}

\subsection{Spine-Wise Pipeline for CNN}
\label{sec:spine_level_pipeline}
\cref{fig:data_dependence}(a) illustrates the spine-wise pipeline in \sysname. 
Spines within a convolution are data-independent of each other, allowing their computations to be performed concurrently. To start the computation of the next layer as early as possible, spines are processed in a specific order, rather than the conventional top-to-bottom, left-to-right sequence. The calculation order is indicated by the arrows in \cref{fig:data_dependence}(a).
For example, we present a timeline showing the computation times of three convolution layers in \cref{fig:data_dependence}(a). In the timeline, layer 3 starts to calculate spine $S_1$ after spine $S_2$ in layer 2 finishes the calculation, as the $S_1$ in layer 3 is data-dependent on $S_1, S_2, S_4, S_5$ in layer 2, which is illustrated by the dark orange regions in \cref{fig:data_dependence}. 
A more general formulation of the control algorithm in \texttt{Output Scheduler} for spine-wise pipeline is provided in \cref{alg:control_algorithm}, where the padded spine is excluded from the input spine. Therefore, the calculation of padded spines is skipped (line 1). 
Then, \sysname generates the position of output spines with the order shown by the blue arrows in \cref{fig:data_dependence} (lines 4-12).
Finally, the calculation of output spines that are data-dependent to padded spines is delayed until the last input valid spine arrives (lines 14-18).

\subsection{Token-Wise Pipeline for Transformer}
\cref{fig:data_dependence}(b) displays the token-wise pipeline in \sysname. Since the data dependence only exists within the same token, \sysname processes spikes token by token, making the token-wise pipeline among SNN layers. In detail, \sysname executes spike operators (listed in \cref{tab:back_operators}) token-wise, triggering the next operator immediately after completing the first token of the current operator.
Since generating a single token in the ssoftmax requires all query and key tokens to be available, \sysname stalls the pipeline to wait for QK spike-attention.

\subsection{Data Storage, Transfer and Management}
\label{subsubsec:data_storage_transfer}
To enable spine/token-wise pipelining, \sysname employs a hierarchical data management strategy: \textbf{1) Intra-core storage.} Partial sums are accumulated in membrane buffers attached to each ST-BIF neuron, serving as local state registers across time-steps. When a neuron fires, the control module reads the membrane and spike tracer and forwards them to the ST-BIF circuit for spike generation and state update. \textbf{2) Inter-core transfer.} Spikes delivered to spines/tokens in the next layer are packed into \emph{flits} and temporarily stored in \texttt{FIFO Queues}, which act as \textit{pipeline registers} between adjacent cores to enable non-blocking, in-order transmission.

\section{SNN Mapping Optimization}
\label{sec:mapping}

As shown in \cref{fig:mapping}, \sysname maps SNN into neural cores through a three-stage mapping algorithm, including partition, mapping, and routing. The mapping algorithm has three targets: 1) minimize the NoC traffic, 2) minimize the required peak bandwidth (\aka RPB), and 3) maximize PE utilization.

\begin{figure}[t]
\centering
\includegraphics[width=\linewidth]{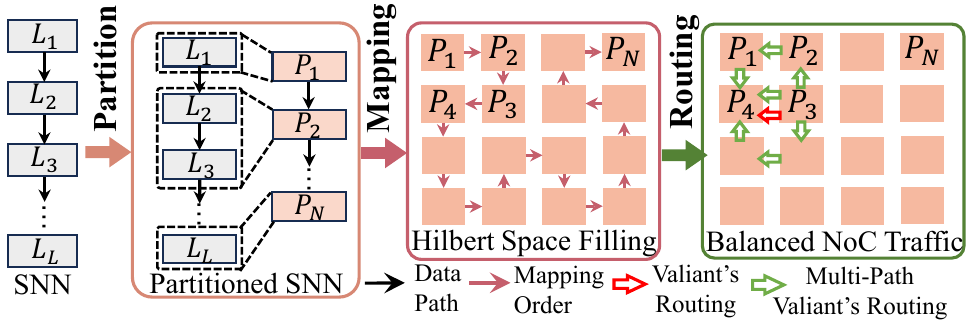}
    \vspace{-1.0em}
    \caption{\textbf{Mapping Procedure in \sysname}. \sysname maps SNN through three stages: partition, mapping, and routing.}
    \label{fig:mapping}
\vspace{-0.5em}
\end{figure}

\textbf{Partition:} In the partition stage, as shown in \cref{fig:mapping}(left), layer-wise partition is preferred in \sysname. The reason is that the communication within SNN layers, such as spike broadcast in tiling strategy (\cref{sec:tiling}) and spike reduction operation between PEs in \texttt{Local Input Reducer}, can be avoided. 
\hl{For mapping a layer across multiple neural cores, we use the MM-sc tiling strategy \cref{fig:SMM}, which column-wise partitions the synaptic weight and membrane matrices across cores.}
To minimize the NoC traffic and maximize the PE utilization, we propose a greedy partition algorithm (\cref{alg:partition_algorithm}).
Firstly, we sort connections $\vc_{ij}$ (line-3). Then, for each connection $\vc_{ij}$, we compare the allocated memory $a = a_i + a_j$ and the number of neuron circuits $d = d_i + d_j$ to the neural core's capacities ($A,D$) (line 5). If within the capacity ($A,D$), we combine the two SNN layers into one partition (lines 6-7). 

\textbf{Mapping:} After partitioning, \sysname applies the Hilbert-curve–based\footnote{Hilbert curve \cite{modified_hilert_curve}: a continuous, fractal space-filling curve that recursively maps a one-dimensional interval onto two-dimensional space.} mapping algorithm from \cite{Mapping_Zheda} to assign partitions to neural cores. As illustrated in the center of \cref{fig:mapping}, the algorithm first generates an initial placement by traversing the Hilbert curve, then models inter-core communication as force potentials and iteratively refines the mapping using a greedy minimization of the total potential.

\textbf{Routing:} After mapping, routing is essential for balancing NoC traffic. 
As shown in \cref{fig:mapping}(right), X–Y routing selects a single path between adjacent neuron cores (\eg, $P_3$ to $P_4$, red hollow arrow), leading to congestion and elevated required peak bandwidth (\aka RPB). 
To mitigate this, we propose a \emph{multi-path routing} algorithm that explores two alternative paths beyond the shortest one (green arrows), bypassing hotspots and enhancing load balance. 
A genetic algorithm is then employed to optimize the transmission probabilities across these paths, further mitigating traffic imbalance.

\begin{algorithm}[t]
\caption{The greedy partition algorithm in \sysname.}
\small
\label{alg:partition_algorithm}  
Input: The required memory $\va$ and \# of neuron circuits of SNN layers $\vd$, neuron core memory $A$, \# of neuron circuits in a neural core $D$, \# of SNN layers $L$, \# of neural cores $N$ and communication traffic $\vc_{ij}$ from $i$-th to $j$-th layer. 

Output: A set list of partitions for SNN layers $\vs$.

sort($\vc$, reverse=True) // sort the communication traffic

\ForEach{$\vc_{ij}$}{
    \If{$\vd_i + \vd_j < D$ and $\va_i + \va_j < A$}{

        $\vd_i$ = $\vd_j$ = $\vd_i + \vd_j$; $\va_i$ = $\va_j$ = $\va_i + \va_j$;

        $\vs_i$ = $\vs_i$ + \{$i,j$\} // add layer $i$,$j$ to the partition.
    }
}
\end{algorithm}

\section{Evaluation}
\subsection{Experimental Setup}
\label{sec:exp_setup}

\subsubsection{Implementation}
We implement the RTL of modules in \sysname and synthesize them with Synopsys Design Compiler using commercial 28nm technology. 
Afterward, we perform post-synthesis simulations with Synopsys VCS to validate the functionality and annotate the toggle rate of the gate-level netlists, where the annotated switch activities are used to estimate the energy consumption with Synopsys PrimeTime PX. 
For memory, the area and energy of SRAM are generated via a commercial memory compiler. 
The off-chip access cost is evaluated using DRAMSim3~\cite{li2020dramsim3} with HBM3.0~\cite{o2014highlights}. 
For network-on-chip, we use DSENT \cite{DSENT} to simulate its energy and latency.
Lastly, we build a cycle-level simulator with each component parameterized from the synthesized results (\ie, area, power, and latency) of ASIC designs. 

\begin{table}[t]
\centering
\caption{\textbf{Specifications of Evaluation Benchmarks}.}
\vspace{-0.5em}
\label{tab:benchmarks}
\renewcommand\arraystretch{0.9}
\resizebox{\columnwidth}{!}{%
\begin{tabular}{cccccccc}
\toprule
\textbf{Work} & \textbf{Topology} & \textbf{Dataset} & \textbf{T.S.}$^\dagger$ & \textbf{\#Ops} & \textbf{\#Sops}$^\ddagger$ & Param. \\ \midrule
W1    & VGG16     & CIFAR10    & 32 & 0.66G & 0.62G & 32.1M \\
W2    & VGG16     & CIFAR100    & 32 & 0.66G & 0.62G & 32.4M \\
W3    & VGG16     & CIFAR10-DVS    & 32 & 1.55G & 2.55G & 32.1M \\
W4    & ResNet18  & ImageNet     & 32 & 3.63G & 3.22G & 11.7M \\
W5    & ResNet34  & ImageNet     & 32 & 7.36G & 9.43G & 21.8M \\
W6    & ResNet50  & ImageNet     & 32 & 8.18G & 10.04G & 25.6M \\
W7    & ViT Small & ImageNet    & 32 & 8.50G & 90.74G & 22.1M \\ 
\multirow{2}{*}{\hl{W8}} 
& \multirow{2}{*}{\hl{YOLOv2}} 
& \hl{COCO2017} 
& \multirow{2}{*}{\hl{32}} 
& \multirow{2}{*}{\hl{18.44G}} 
& \multirow{2}{*}{\hl{37.63G}} 
& \multirow{2}{*}{\hl{52.8M}} \\
& & \hl{VOC2017} & & & & \\
\hl{W9}    & \hl{ResNet101} & \hl{ImageNet} & \hl{32} & \hl{15.60G} & \hl{19.61G} & \hl{44.5M} \\ 

\bottomrule
\multicolumn{7}{l}{$\dagger$T.S. denotes allowed maximum time-steps. $\ddagger$Sop denotes synaptic operation.}                      
\end{tabular}%
}
\vspace{-1.0em}
\end{table}

\begin{table}[t]
\centering
\caption{\textbf{Hardware Specifications of \sysname}.}
\vspace{-1.0em}
\label{tab:hardware_analysis}
\renewcommand\arraystretch{0.9}
\resizebox{\columnwidth}{!}{%
\begin{tabular}{rcccc}
\toprule
\multicolumn{1}{c}{\textbf{\begin{tabular}[c]{@{}c@{}}Component\\ Name\end{tabular}}} &
  \textbf{Metric} &
  \textbf{Spec.} &
  \textbf{\begin{tabular}[c]{@{}c@{}}Power (\bm{$\mu$}W)/\\ Percentage\end{tabular}} &
  \textbf{\begin{tabular}[c]{@{}c@{}}Area (mm\textsuperscript{2})/\\ Percentage\end{tabular}} \\ \midrule
\multicolumn{5}{c}{\{Process Elements $\times$ 4\} in Single Neural Core}                                \\ \midrule
\texttt{Weight Memory}                & size        & 4$\times$102.4 KB & 715.0/31.2\%   & 0.487/17.49\%     \\
\texttt{Membrane Memory}                     & size        & 4$\times$307.2 KB & 96.1/4.2\%    & 1.460/52.44\%     \\
\texttt{Spike Tracer Memory}                 & size        & 4$\times$102.4 KB  & 13.6/0.6\%    & 0.487/17.49\%      \\
\texttt{FireComponent}                & count       & 4$\times$128   & 84.7/3.7\%    & 0.0189/0.7\%     \\
\texttt{16-input Adder Tree}          & count       & 4$\times$128   & 1191.4/52.0\%  & 0.140/5.02\%      \\
\multicolumn{1}{c}{Sub-total}    & -           & -              & 2103.3/91.8\%  & 2.59/93.03\%      \\ \midrule
\multicolumn{5}{c}{\{Router\} in Single Neural Core}                                                     \\ \midrule
\texttt{SLayerNorm Unit}                       & count       & 1              & 33.7/1.5\%    & 0.091/3.27\%     \\
\texttt{SSoftmax Unit}                       & count       & 1              & 43.1/1.9\%    & 0.096/3.45\%     \\
\texttt{FIFO Queue}                   & size        & 4$\times$512 B        & 91.6/4.0\%    & 0.0013/0.047\%   \\
\texttt{Flit Generator}               & count       & 1              & 2.9/0.1\%     & 0.0011/0.040\%   \\
\texttt{Crossbar Switch}              & count       & 1              & 16.4/0.7\%    & 0.0017/0.061\%   \\
\texttt{Others}                       & -           & -              & 0.2/0.0\%     & 0.00015/0.0054\% \\
\multicolumn{1}{c}{Sub-total}    & -           & -              & 187.9/8.2\%   & 0.19/6.97\%      \\ \midrule
\multicolumn{1}{c}{\sysname Chip} & \#Tiles & $6\times 6$    & \textbf{82490.0}/100\% & \textbf{100.23}/100\%            \\ \bottomrule
\end{tabular}%
}
\vspace{-1.5em}
\end{table}

\begin{table*}[t]
\centering
\caption{\textbf{Comparison with SNN accelerators.}}
\vspace{-0.5em}
\label{tab:HardwareCompare}
\setlength{\tabcolsep}{2.5pt}
\renewcommand\arraystretch{0.9}
\resizebox{\linewidth}{!}{
\begin{tabular}{@{}rcccccccccccc@{}}
\toprule
\multicolumn{1}{l}{} 
& \textbf{SpinalFlow\hspace{-0.5pt}\cite{narayanan2020spinalflow}} 
& \textbf{Prosperity\hspace{-0.5pt}\cite{prosperity}} 
& \textbf{SASAP\hspace{-0.5pt}\cite{SASAP}} 
& \textbf{Phi\hspace{-0.5pt}\cite{phi}} 
& \textbf{C-DNN\hspace{-0.5pt}\cite{C-DNN}}
& \textbf{MorphIC\hspace{-0.5pt}\cite{morphic}} 
& \textbf{TrueNorth\hspace{-0.5pt}\cite{akopyan2015truenorth}} 
& \textbf{Darwin\hspace{-0.5pt}\cite{darwin3}} 
& \textbf{PAICORE\hspace{-0.5pt}\cite{PAICORE}} 
& \multicolumn{3}{c}{\textbf{\sysname}} \\ 
\midrule

Technology 
& 28 nm & 28 nm & 40nm & 28nm & 28nm 
& 65nm & 65nm & 22nm & 28nm 
& \multicolumn{3}{c}{28nm} \\

Voltage(V) 
& 0.9 & n/a & 0.56-1.1 & n/a & 0.7-1.1 
& 0.8-1.2 & 0.7-1.04 & 0.8 & 0.675 
& \multicolumn{3}{c}{0.9} \\

Freq.(MHz) 
& 200 & 500 & 50-200 & 500 & 50-200 
& 55-210 & 0.001$\triangle$ & 333 & 168 
& \multicolumn{3}{c}{200-500} \\

\hl{SRAM Only} 
& No & No & No & No & No 
& Yes & Yes & Yes & Yes 
& \multicolumn{3}{c}{Yes} \\ 

Core Number 
& 1 & 1 & 2 & 1 & 64 
& 4 & 4096 & 575 & 1024 
& \multicolumn{3}{c}{36} \\

Area(mm$^2$) 
& 2.09 & 0.529 & 2.69 & 0.662 & 20.25 
& 2.86 & 430 & 358.53 & 537.98 
& \multicolumn{3}{c}{100.23} \\

ALU per PE 
& 8-b Add,Cmp 
& 8-b Add 
& 8-b Add,Cmp 
& 8-b Add 
& \begin{tabular}[c]{@{}c@{}}1-16 bit\\ MAC,Add,CMP\end{tabular}
& 1-b Add,Cmp 
& 8-b Add,Cmp 
& \begin{tabular}[c]{@{}c@{}}1/2/4/8/16-bit\\ Add, Cmp\end{tabular} 
& 1/8-b Add,Cmp 
& \multicolumn{3}{c}{\begin{tabular}[c]{@{}c@{}}8-b Add,\\ Cmp, Shift\end{tabular}} \\

Memory 
& 585 KB & 136 KB & n/a & 240 KB & 552 KB 
& 288 KB & 51 MB & \textgreater{}23.44 MB & 121.87 MB 
& \multicolumn{3}{c}{72 MB} \\

Scheduling 
& NoPipe & NoPipe & NoPipe & NoPipe & NoPipe 
& Layer & Layer & Layer & Layer 
& \multicolumn{3}{c}{Spine/Token} \\

GOPS 
& 684.5 & 390.1 & 72.5\textsuperscript{1} & 242.8 & 842.8\textsuperscript{3}
& 0.42 & 58.0 & 66.8 & 1421.6\textsuperscript{2} 
& 1982.9 & \textbf{4135.4} & 2315.1 \\

TOPS/W 
& 4.22 & 0.299 & 42.8\textsuperscript{1} & 0.286 & 24.5\textsuperscript{3}
& 0.29 & 0.400 & 0.18 & 1.156\textsuperscript{2} 
& 20.89 & \textbf{25.55} & 5.10 \\

pJ/Sops$\ddagger$ 
& n/a & n/a & 0.078\textsuperscript{1} & n/a & 1.1\textsuperscript{3}
& 51 & n/a & 5.47 & 0.865\textsuperscript{2} 
& 0.051 & 0.032 & \textbf{0.020} \\

\hl{GOPS/mm$^2$} 
& 327.5 & 737.4 & 27.00 & 366.8 & 41.62
& 0.147 & 0.134 & 0.186 & 2.642 
& 19.78 & 41.26 & 23.10 \\

Network 
& ResNet34 & VGG16 & Spikformer & VGG16 & ResNet50
& n/a & n/a & VGG16 & ResNet50 
& VGG16 & ResNet50* & VIT-S \\

Accuracy (\%) 
& IN@65.5 & CF10@92.3 & IN@77.1 & CF10@91.1 & IN@75.2
& MNIST@97.8 & n/a & CF10@90.2 & IN@77.1 
& \textbf{CF10@92.3} & IN@75.6 & \textbf{IN@79.1} \\

Elastic Infer.
& \xmark & \xmark & \xmark & \xmark & \xmark
& \cmark & \cmark & \cmark & \cmark 
& \cmark & \cmark & \cmark \\

\bottomrule
\multicolumn{13}{l}{$\star$: The frequency of global tick in TrueNorth \cite{akopyan2015truenorth} is 1kHZ. IN denotes the ImageNet dataset and CF10 denotes the CIFAR10 dataset.} \\ 
\multicolumn{13}{l}{*: after prune. $\ddagger$: (energy/Frame)/(\# of pre-synaptic and post-synaptic Spikes). \textsuperscript{1}: evaluated with 0.56 V and 50 MHZ. \textsuperscript{2}: evaluated with 4-bit synaptic weight. \textsuperscript{3}: evaluated with 0.7 V and 50 MHZ.} \\ 
\end{tabular}}
\vspace{-1.0em}
\end{table*}

\begin{table*}[t]
\centering
\caption{\textbf{Comparison of \sysname \wrt QANN Accelerators}. All designs are evaluated with the voltage of 0.9 V.}
\vspace{-0.5em}
\label{tab:QANN_comparison}
\setlength{\tabcolsep}{2.5pt}
\renewcommand\arraystretch{0.9}
\resizebox{\linewidth}{!}{%
\begin{tabular}{rcccccccccccc}
\toprule
 & \textbf{Eyeriss$^\dagger$\hspace{-0.5pt}\cite{7738524}} & \textbf{Eyeriss v2$^\ddagger$} \cite{EyerissV2} & \textbf{ANT\hspace{-0.5pt}\cite{ANT}} & \textbf{S-CONV\hspace{-0.5pt}\cite{S-CONV}} & \textbf{AIOQAB\hspace{-0.5pt}\cite{AICAS2024}} & \textbf{Sanger\hspace{-0.5pt}\cite{Sanger}} & \textbf{ViTALiTy\hspace{-0.5pt}\cite{ViTALiTy}} & \hl{\textbf{AEC-CIM\hspace{-0.5pt}\cite{AEC-CIM}}} & \textbf{LLH-CIM\hspace{-0.5pt}\cite{LLH-CIM}} & \multicolumn{3}{c}{\textbf{\sysname}} \\ \midrule
Implementation & Digital & Digital & Digital & Digital & Digital & Digital & Digital & Digital CIM & Analog CIM & \multicolumn{3}{c}{Digital} \\
Technology & 28nm & 28nm* & 28nm & 28nm & 28nm & 28nm & 28nm & 28nm & 22nm & \multicolumn{3}{c}{28nm} \\
Frequency & 200MHz & 200MHz & 200MHz & 400MHz & 500MHz & 667MHz & 500MHz & n/a & 244MHz & \multicolumn{3}{c}{200-500MHz} \\
Area(mm\textsuperscript{2}) & 2.969 & 1.536* & 4.527 & 2.69 & 0.592 & 5.194 & 5.223 & 0.468 & 0.119 & \multicolumn{3}{c}{100.23} \\
ALU per PE & 8-b MAC & 8-b MAC & 4.8-b MAC & 8-b MAC & 4-b MAC & 16-b MAC & 16-b MAC & 8-b MAC & 8-b MAC & \multicolumn{3}{c}{8-b Add,Cmp,Shift} \\
Network & ResNet50 & ResNet50 & ResNet50 & ResNet34 & ViT-S & ViT-S & ViT-S & n/a & ResNet18 & ResNet18 & ResNet50 & ViT-S \\
GOP/s$^{**}$ & 40.26 & 153.6 & 1210.06 & 741.93 & 132.25 & 615.50 & 2057.61 & 213.4 & 62.4 & 1347.84 & \textbf{4135.42} & 2315.14 \\
TOPS/W$^{**}$ & 0.766 & 2.336* & 1.880 & 4.907 & 1.789 & 0.365 & 1.25 & 22.75 & 20.7 & 29.87 & \textbf{25.55} & 5.10 \\
GOPS/mm\textsuperscript{2} & 13.56 & 100.01* & 264.78 & 275.81 & 223.39 & 118.50 & 393.95 & 456.00 & 524.37 & 13.45 & 41.26 & 23.10 \\
Accuracy(\%) & IN@75.97 & IN@75.6 & IN@75.08 & IN@71.8 & IN@78.5 & IN@79.2 & IN@79.5 & n/a & IN@69.25 & IN@69.5 & IN@75.6 & IN@\textbf{79.1}$^\triangle$ \\
Elastic Inference & \xmark & \xmark & \xmark & \xmark & \xmark & \xmark & \xmark & \xmark & \xmark & \cmark & \cmark & \cmark \\ \bottomrule
\multicolumn{13}{l}{$\dagger$: performance of Eyeriss is reproduced from Accelergy \cite{lee2020reconfigurable}; $^\triangle$: ViT-S on \sysname is trained by ourselves, while other accuracies are taken from original work. IN is short for ImageNet.} \\
\multicolumn{13}{l}{$\ddagger$: data from Eyeriss V2 \cite{EyerissV2}; *: the performance is scaled to 28nm; $^{**}$: 1 MAC=2 OP, \#time-step Sop=2 OP, TOPS/W = \#OP of network / Latency per frame, \sysname is measured at 200MHz. } \\
\end{tabular}}
\vspace{-1.0em}
\end{table*}

\subsubsection{Benchmarks}
The evaluation benchmarks are listed in \cref{tab:benchmarks}. \hl{For the classification task}, benchmarks are SNNs converted from CNN (\ie, VGG16 \cite{Simonyan2014VeryDC} and \hl{ResNet18/34/50/101} \cite{resnet}) and Transformer (\ie, ViT Small \cite{visionTransformer}) using datasets of CIFAR-10/100 \cite{alex2009cifar}, CIFAR10-DVS \cite{li2017cifar10dvs}, and ImageNet \cite{deng2009imagenet}. 
\hl{For the detection task, benchmarks are SNNs converted from YOLOv2 with ResNet34 as backbone on COCO2017 \cite{lin2014microsoft} and VOC2007 \cite{everingham2010pascal} datasets.}
Note that all SNNs in \cref{tab:benchmarks} use 4-bit quantized weights, and all the benchmark latencies are obtained after the accuracy converges in elastic inference. Evaluated SNNs are generated following SpikeZIP-TF \cite{spikeziptf2024}.

\subsubsection{Baselines}
We compare ELSA with three categories of baselines to comprehensively demonstrate improvements:

\begin{itemize}[leftmargin=*]
\item \textbf{Elastic SNN accelerators.} 
We predominantly compare ELSA with SNN accelerators that support elastic inference, including TrueNorth\cite{akopyan2015truenorth}, Darwin\cite{darwin3}, MorphIC\cite{morphic}, and PAICORE\cite{PAICORE}, \textit{to highlight the benefits of architectural innovations of ELSA.}
These accelerators support TBT execution and elastic inference as discussed in \cref{sec:introduction}, and exploit common optimizations such as near-SRAM execution, addition-only computation, and event-driven sparsity. 
\textit{ELSA shares the same foundations, enabling fair comparisons.}

\item \textbf{Non-elastic SNN accelerators.} 
We further compare ELSA with SNN accelerators based on LBL execution without elastic inference capability, including Phi \cite{phi}, SpinalFlow \cite{narayanan2020spinalflow}, SASAP \cite{SASAP}, Prosperity \cite{prosperity}, and C-DNN \cite{C-DNN}, to provide a more comprehensive scope of comparison.

\item \textbf{QANN accelerators.} 
To highlight the benefits of SNN features (\ie, event-driven sparsity and addition-only computation), we compare ELSA with multiple state-of-the-art QANN accelerators spanning digital designs (Eyeriss\cite{7738524}, Eyeriss v2\cite{EyerissV2}, ANT\cite{ANT}, S-CONV\cite{S-CONV}, AIOQAB\cite{AICAS2024}, Sanger\cite{Sanger}, and  ViTALiTy \cite{ViTALiTy}), digital \quad in-memory design (AEC-CIM\cite{AEC-CIM}), and analog in-memory design (LLH-CIM\cite{LLH-CIM}).
We also compare ELSA with commercial accelerators, including Jetson AGX Orin 64GB \cite{Jetson_AGX_Orin}, Nvidia A100 GPU \cite{Nvidia_A100}, TPU v4 \cite{TPUV4}, and Groq \cite{Groq}, which have comparable chip area to \sysname. 

\end{itemize}

\subsubsection{Metrics Modeling}
 We model the metrics of competing designs through two steps. 1) For latency and energy values reported in the original papers, we directly use those values. 2) For cases where such metrics are not provided, we estimate them using the reported peak throughput and peak energy efficiency, to enable a relatively fair comparison.

\subsubsection{Early Termination} \sysname conducts the early termination by a confidence-based method for classification and detection \cite{SEENN, OEDSNN, SpikeCP} to reduce latency while maintaining task accuracy. On the classification task, we use the maximum class probability as the confidence score and terminate inference at intermediate time-steps once the confidence exceeds a predefined threshold. 
For detection tasks, we use the objectness score produced by the detector (\eg, YOLO \cite{redmon2016you}) as the confidence for early termination.

\subsection{Breakdown of Components in \sysname}

\textbf{Power and Area Breakdown.}
\cref{tab:hardware_analysis} lists the power and area cost of the components used to build \sysname. 
\sysname organizes 6$\times$6 neural cores in a 2D-mesh, and each neural core contains 4 PEs and 1 router. The PEs and routers consume 93.03\% and 6.97\% of the total area, respectively. 
The PE area is dominated by various memories, storing weight, membrane, and spike tracer of each neuron, which takes 93.97\% of the PE area and 93.03\% of the entire \sysname. 
The PE area could be further optimized by replacing massive SRAM with other on-chip embedded memories (\eg, eDRAM \cite{matick2005logic, mittal2014survey}) via advanced integration technology.
The router is mostly occupied by \texttt{SSoftmax Unit} and \texttt{SLayerNorm Unit} (\ie, 6.72\% of \sysname). The reason is that \texttt{SSoftmax Unit} and \texttt{SLayerNorm Unit} contain ST-BIF neuron circuits and memories to store spike tracer and membrane.
The power of \sysname is mainly consumed by \texttt{adder tree} (52\%) and \texttt{weight memory} (31.2\%) since SNN inference is dominated by spike-driven addition and weight access. 

\begin{figure}[t]
\centering
    \includegraphics[width=\linewidth]{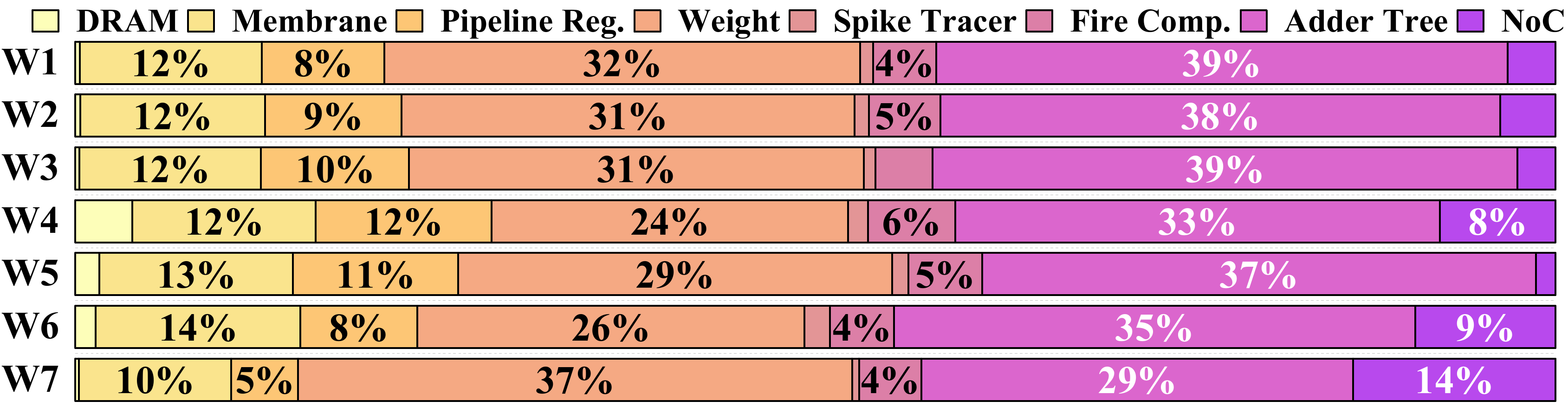}
    \vspace{-1.5em}
    \caption{\textbf{Energy breakdown of \sysname}  on the benchmark W1-7 (\cref{tab:benchmarks}). Fire Comp. is short for fire component. The Pipeline Register Energy is consumed by \texttt{FIFO Queue}. }
    \label{fig:energy_breakdown}
    \vspace{-1.5em}
\end{figure}

\textbf{Energy Breakdown} of \sysname is shown in
\cref{fig:energy_breakdown}, where \texttt{adder tree} consumes most of the energy ($29\% \sim 39\%$) at most of benchmarks. The second-highest energy consumption comes from memory access, including \texttt{FIFO Queue} and \texttt{Membrane, Weight, Spike-Tracer Buffer}.
Thanks to the dataflow and near-memory computing design in \sysname, the off-chip DRAM access is negligible, as only the inputs of SNN are loaded from DRAM.

\subsection{Comparison with SNN Accelerators}
\label{sec:SNN_comparison}
Comprehensive comparisons with prior SNN accelerators are in \cref{tab:HardwareCompare}. 
Further comparisons across six specific benchmarks are shown in \cref{SNN_benchmarks}, where ELSA prominently outperforms prior SNN accelerators on various tasks.

\textbf{Compared to elastic SNN accelerators} (TrueNorth\cite{akopyan2015truenorth}, Darwin\cite{darwin3}, MorphIC\cite{morphic}, and PAICORE\cite{PAICORE}), ELSA achieves the highest throughput and energy efficiency, highlighting the effectiveness of proposed architectural innovations.
Specifically, compared to TrueNorth\cite{akopyan2015truenorth}, ELSA breaks computation dependencies via mini-batch spiking Gustavson-product dataflow and BAER, thereby eliminating global synchronization barriers (\ie, 1 ms global tick in TrueNorth\cite{akopyan2015truenorth}) and significantly improving throughput from 58.0 GOPS to 4135.4 GOPS.
Compared to the SOTA accelerator PAICORE \cite{PAICORE} in various benchmarks (\cref{SNN_benchmarks}), ELSA achieves $27.4\times$ geomean energy-saving from the mini-batch spiking Gustavson-product dataflow, and $1.65\times$ geomean speedup from the spine/token-level fine-grained pipeline, respectively.
Compared to Darwin\cite{darwin3}, ELSA consistently achieves better performance, but falls short in bit-width precision flexibility, as Darwin\cite{darwin3} supports 1/2/4/8/16-bit computation for broader programmability.
Note that elastic SNN accelerators generally show lower area efficiency than non-elastic accelerators (\eg, Prosperity\cite{prosperity}), as all weights and membrane states are stored on-chip.
Nevertheless, \sysname~achieves the highest area efficiency among elastic SNN accelerators, thanks to the spine/token-level pipeline improving the throughput (\cref{fig:ablution_study}).

\textbf{Compared to non-elastic SNN accelerators} (C-DNN\cite{C-DNN}, SpinalFlow\cite{narayanan2020spinalflow}, Prosperity\cite{prosperity}, SASAP\cite{SASAP}, and Phi\cite{phi}) without elastic inference capability, \sysname~achieves the highest throughput (4.9$\times$ higher than the SOTA accelerator C-DNN\cite{C-DNN}), since \sysname~has larger on-chip hardware resources and leverages spine/token-level pipeline to reduce end-to-end latency (\cref{fig:pipeline_comparison}).
The energy efficiency is also improved by mini-batch spiking Gustavson product that reduces the memory access (\cref{fig:product_dataflow}).
However, the gain is marginal (24.5 TOPS/W in C-DNN vs. 25.6 TOPS/W in \sysname), as C-DNN (LBL-based) avoids SRAM storage for membrane states.

\begin{figure}[t]
\centering
    \includegraphics[width=\linewidth]{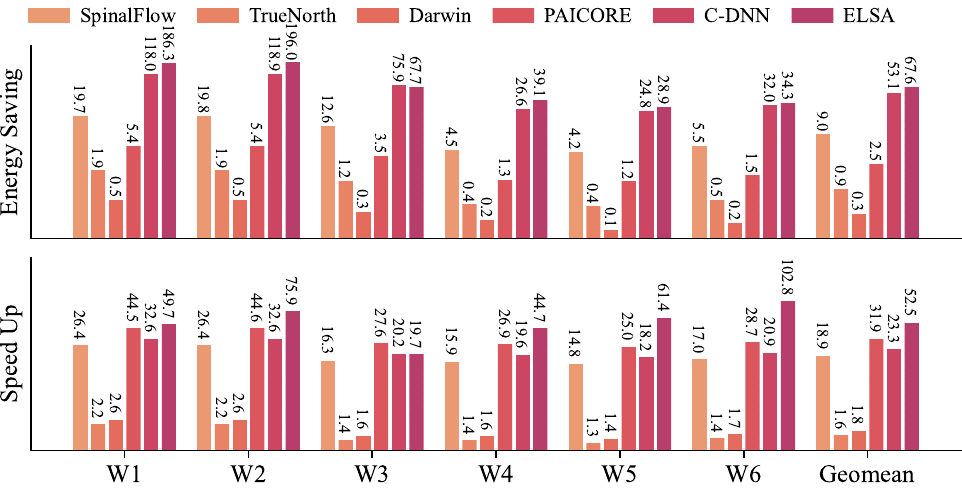}
    \vspace{-1.5em}
    \caption{\textbf{Energy and latency comparison of SNN accelerators.} Statistics are normalized \wrt Eyeriss \cite{7738524}.}
    \label{SNN_benchmarks}
\vspace{-0.5em}
\end{figure}

\subsection{Comparison with QANN Accelerators}
\label{sec:compare_with_QANN}

Since SNNs adopted in \sysname~are converted from QANN models with the same accuracy, we compare \sysname~with existing QANN accelerators to demonstrate that the intrinsic advantages of SNNs are effectively exploited under an equal-accuracy setting.
\cref{tab:QANN_comparison} presents the throughput and energy efficiency of \sysname~operating at 200MHz, targeting low-power scenarios. 
Compared to the SOTA ResNet50 accelerator ANT~\cite{ANT}, \sysname~delivers $3.4\times$ higher throughput and $13.6\times$ better efficiency.
Compared to the SOTA ViT-S accelerator ViTALiTY~\cite{ViTALiTy}, \sysname~achieves $2.8\times$ throughput (\ie, 5787 GOP/s) and $4.1\times$ efficiency improvements at an aligned 500MHz frequency.
\hl{Compared to the digital-CIM design (AEC-CIM \cite{AEC-CIM}) and analog-CIM design (LLH-CIM \cite{LLH-CIM}), \sysname~achieves $1.31\times$ and $1.44\times$ better energy efficiency.}

\begin{figure}[t]
\centering
    \includegraphics[width=\linewidth]{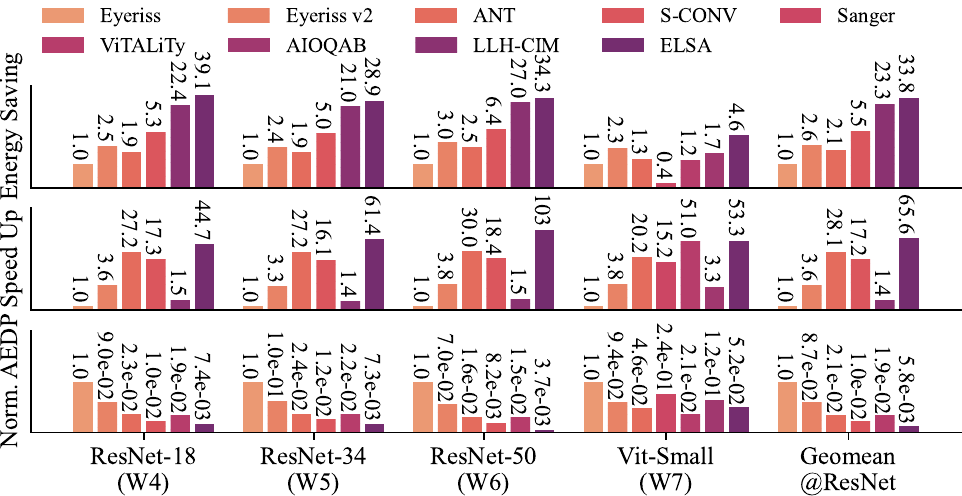}
    \vspace{-1.5em}
    \caption{\textbf{Energy, latency, AEDP comparison with QANN accelerators}. Statistics are normalized \wrt Eyeriss \cite{7738524}.}
    \label{fig:QANN_comparsion}
\vspace{-0.5em}
\end{figure}

As shown in \cref{fig:QANN_comparsion}, we compare \sysname with prior QANN accelerators across multiple benchmarks in terms of latency, energy, and AEDP. For a fair system-level comparison, we include an additional 16 MB eDRAM \cite{huang2011high} (4.76 mm\textsuperscript{2}) for weight storage. Overall, \sysname achieves the best speedup and energy efficiency on most workloads. 
Note that, on ViT-S, \sysname yields a higher AEDP than ViTALiTy \cite{ViTALiTy}, as the $9.0\times$ larger SOP count (\cref{tab:benchmarks}) reduces both GOPS and TOPS/W.

\subsection{Comparison with Large Chip}
\label{sec;large_on_chip_accelerator}
\cref{tab:GPU_comparison} compares \sysname with QANN accelerators with large on-chip memory and die area. Thanks to the lossless conversion algorithm, the task accuracies of QANN in GPUs/Groq/TPU, and SNN in \sysname are the same (\ie 75.6\% accuracy on ImageNet-1K with ResNet50). Compared to the edge GPU Jetson AGX Orin, \sysname achieves 144.4$\times$ energy efficiency (TOPS/W) improvement and 49.9\% chip area (mm$^2$) reduction, showing competing performance in the edge application. Compared to high-performance GPU A100 and dataflow architectures (TPU V4 and Groq), \sysname~has lower throughput (TOPS) since these accelerators have better chip technology ($<$14 nm), higher frequency ($>$900 MHz), and larger area ($>$700 mm$^2$). Thanks to multi-level optimizations and the inherently low energy consumption of the SNN algorithm, \sysname achieves the highest energy efficiency (8.2 $\times$ improvement compared to Groq) among them.
\begin{table}[t]
\caption{Comparison of Large Chips running ResNet50.}
\vspace{-0.5em}
\label{tab:GPU_comparison}
\renewcommand\arraystretch{0.85}
\resizebox{\linewidth}{!}{
\begin{tabular}{@{}lccccc@{}}
\toprule
 & \multicolumn{1}{l}{Jetson AGX Orin} & A100 & TPU V4 & Groq & ELSA  \\ \midrule
Implementation & Digital & Digital & Digital & Digital & Digital \\
Dataflow Arch.? & No & No & Yes & Yes & Yes \\
Technology & 8nm & 7nm & 7nm & 14nm & 28nm \\
On-chip Memory & 4.25MB & 40MB & 170MB & 230MB & 72MB \\
Frequency(MHz) & 1300 & 1095 & 1050 & 900.0 & 200.0 \\
Area(mm$^2$) & 200.0 & 826.0 & 700.0 & 720.0 & 100.2 \\
TOPS & 10.65 & 624.0 & 275.0 & 750.0 & 4.135 \\
TOPS/W & 0.177 & 1.560 & 1.432 & 3.125 & \textbf{25.55} \\
GOPS/mm$^2$ & 50.33 & 755.4 & 392.8 & 1041.7 & 41.26 \\
\bottomrule
\end{tabular}}
\end{table}

\subsection{Accuracy and Mismatch Analysis for Elastic Inference}

\begin{table}[t]
\caption{\hl{Accuracy of ANN, QANN, SNN, and SNN with early termination (E.T.) on ImageNet, and the latency reduction achieved by early termination (SNN+E.T.) relative to the SNN baseline on CNN and Transformer benchmarks. }}

\label{tab:accuracy_latency}
\resizebox{\linewidth}{!}{
\begin{tabular}{@{}cccccc@{}}
\toprule
\multirow{2}{*}{Method} & \multicolumn{4}{c}{Accuracy} & \multirow{2}{*}{\begin{tabular}[c]{@{}c@{}}Latency \\ Reduction\end{tabular}} \\ \cmidrule(lr){2-5}
 & ANN & QANN & SNN & SNN+E.T. &  \\ \midrule
ResNet18 & 69.61\% & 67.85\% & 67.85\% & 67.79\%/64.38\% & 22.6\%/31.0\% \\
ResNet34 & 74.52\% & 71.54\% & 71.54\% & 71.43\%/68.59\% & 26.1\%/39.1\% \\
ResNet50 & 78.17\% & 75.60\% & 75.60\% & 75.52\%/71.14\% & 16.6\%/19.3\% \\
ViT Small & 81.39\% & 79.07\% & 79.07\% & 78.98\%/76.24\% & 22.3\%/33.1\% \\ \bottomrule
\end{tabular}}
\vspace{-1.0em}
\end{table}

\hl{\textbf{Accuracy analysis} for elastic inference is in \cref{tab:accuracy_latency}, where we provide the accuracies of ANN, QANN, SNN, and SNN with elastic inference. 
QANN Accuracy degradation (\eg, from 78.17\% to 75.60\% on ResNet-50) is common due to quantization and has been widely reported in prior QANN accelerators \cite{ANT, BitFusion, AICAS2024}.
Since the SNN, comprised of ST-BIF neurons, is equivalent to QANN (\cref{sec:background_STBIF}), the accuracies of QANNs and SNNs in ELSA are identical.
With early termination in elastic inference (SNN+E.T.), it achieves an average 21.9\% latency reduction with negligible accuracy loss ($<$ 0.2\% in all benchmarks). With an aggressive confidence threshold choice, ELSA achieves 30.6\% latency reduction with mild accuracy degradation ($<$3.3\%). }

\hl{\textbf{Mismatch analysis} on COCO2017 with YOLOv2 is provided in \cref{fig:mismatch_in_coco}. \textit{If an early-terminated detection has the same class and an IoU (Intersection over Union) greater than 0.5 with the corresponding final detection, we consider it a match.} The definition of confidence and termination criterion is introduced in the experimental setup. As shown in the \cref{fig:mismatch_in_coco} (left), with the increasing confidence threshold, the mismatch rate decreases while the average latency increases. A sweet point is 0.2 confidence, where the match rate is 94.9\% while achieving 45.4\% geometric-mean latency reduction (1.83$\times$ speedup). Importantly, the outputs of elastic inference are progressively refined as computation proceeds. Therefore, with longer inference time, the mismatch rate is reduced to zero. 
We also provide the latency breakdown for the first-correct-response sample at the sweet point (confidence threshold = 0.2) in \cref{fig:mismatch_in_coco} (right). The earliest first-correct-response is 1.19 ms, achieving 2.76$\times$ speedup compared to the full inference, demonstrating that ELSA is well-suited for the latency-critical applications, such as autonomous driving.
}

\subsection{Significance Analysis for Elastic Inference}

\hl{We further analyze the impact of object significance, defined as the ratio between the bounding box area and the image area, in \cref{fig:bbox_in_coco}. As objects become more prominent (area ratio increasing from 0.05 to 0.85 on VOC2007 and from 0.01 to 0.1 on COCO2017), detection terminates earlier. The latency decreases from 2.38 ms to 1.88 ms on VOC2007 and from 1.73 ms to 1.64 ms on COCO2017, indicating that ELSA responds faster to more salient objects. Meanwhile, the mismatch rate remains below 8\% across all object size ranges.}

\begin{figure}[t]
\centering
    \includegraphics[width=\linewidth]{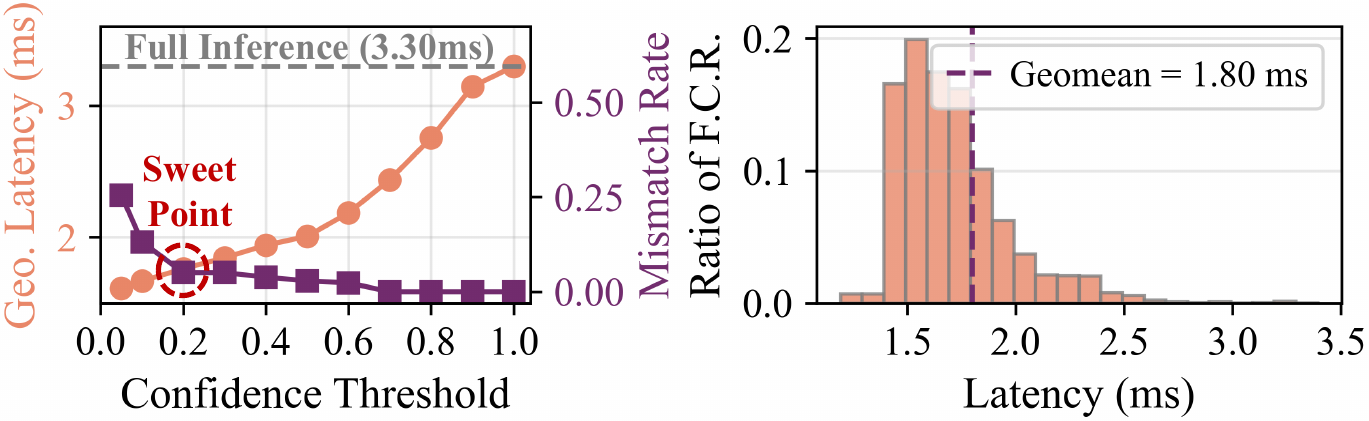}
    \caption{\hl{Mismatch rate (\%) and latency (ms) with different confidence thresholds (left) and latency breakdown (right) under sweet point on COCO2017 dataset with YOLOv2. ``F.C.R." is first-correct-response.}}
    \label{fig:mismatch_in_coco}
    \vspace{-1.0em}
\end{figure}

\begin{figure}[t]
\centering
\includegraphics[width=\linewidth]{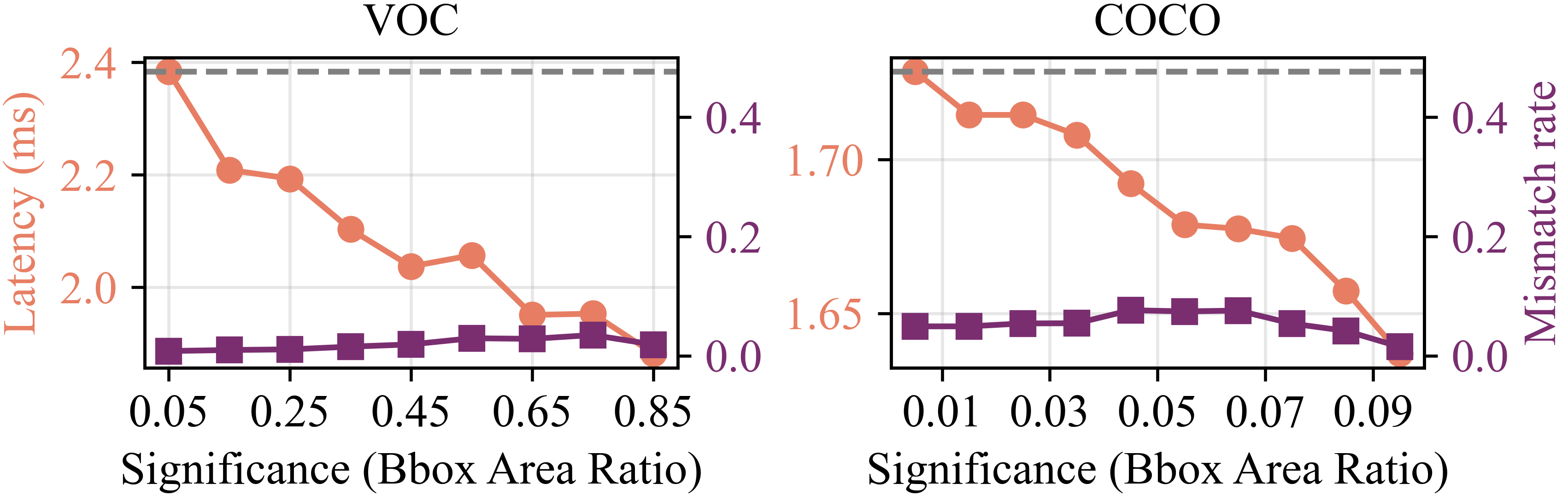}
    \caption{\hl{Latency v.s. Significance (area ratio of bounding box prediction) in VOC2007 \cite{everingham2010pascal} and COCO2017 \cite{lin2014microsoft}. More salient objects with larger area ratios tend to terminate earlier. }}
    \label{fig:bbox_in_coco}
    \vspace{-2.0em}
\end{figure}

\subsection{Network-on-Chip Comparison on ImageNet}
\label{sec:NOC_comparison}
We compare the network-on-chip (NoC) traffic and energy of \sysname\ with other multi-core SNN accelerators, including MorphIC \cite{morphic} and TrueNorth \cite{akopyan2015truenorth}, as summarized in \cref{tab:NoC_traffic_energy_comparison}.
For fairness, all NoCs are configured identically (6×6 2D mesh, same bandwidth and flit buffer size) and modified to support slayernorm, ssoftmax, and im2col for the benchmarks in \cref{tab:benchmarks}.
Benefiting from the bundled AER and multi-path routing, \sysname\ achieves the lowest NoC traffic and energy across all benchmarks, with 20.5\% traffic and 24.3\% energy reductions over TrueNorth on average.

\begin{table}[t]
\caption{NoC traffic–energy comparisons.}
\resizebox{\columnwidth}{!}{%
\begin{tabular}{@{}ccccccc@{}}
\toprule
 & \multicolumn{2}{c}{\textbf{Morphic}} & \multicolumn{2}{c}{\textbf{TrueNorth}} & \multicolumn{2}{c}{\textbf{ELSA}} \\ \midrule
 & Traffic & Energy & Traffic & Energy & Traffic & Energy \\ \midrule
ResNet18 & 15.5MB & 3.24µJ & 12.9MB & 3.22µJ & 11.2MB & 2.63µJ \\
ResNet34 & 29.9MB & 6.81µJ & 26.7MB & 5.69µJ & 21.0MB & 5.05µJ \\
ResNet50 & 92.8MB & 32.1µJ & 72.3MB & 30.5µJ & 52.6MB & 19.3µJ \\
ViT Small & 994.8MB & 0.33mJ & 701.6MB & 0.26mJ & 560.3MB & 0.18mJ \\ \bottomrule
\end{tabular}
\label{tab:NoC_traffic_energy_comparison}
}
\vspace{-1em}
\end{table}

\subsection{Elastic Inference v.s. Spine/Token-wise Pipeline}

\begin{figure}[t]
    \centering
    \includegraphics[width=\linewidth]{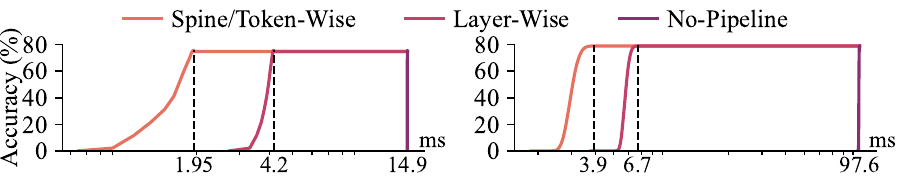}
    \vspace{-1.5em}
\caption{Elastic Inference in \sysname~using three different pipeline schedulings in ResNet50 (Left) and ViT Small(Right). X-axis: latency (ms); Y-axis: top-1 accuracy (\%).}
    \label{fig:VSA_pipeline_elastic}
    \vspace{-1.0em}
\end{figure}

Our spine/token-wise pipeline produces outputs at fine granularity. 
Each token and spine can exit independently once confidence is high. 
This design aligns perfectly with elastic inference. 
As shown in \cref{fig:VSA_pipeline_elastic}, compared to the other coarse-grained pipeline (no pipeline or layer-wise pipeline) used in prior SNN accelerators \cite{akopyan2015truenorth, darwin3, yin2024loas}, the accuracy-latency curve of spine/token-wise pipeline shifts leftward, showing a faster response.
As a result, spine/token-wise pipeline achieves an average $2.0\times$ earlier on ViT-S and $2.4\times$ speedup on ResNet50 compared to other pipelines at the same accuracy. 
This shows that spine/token-wise pipeline enables lower-latency elastic inference, critical for real-time applications.

\begin{figure}[t]
\centering
    \includegraphics[width=\linewidth]{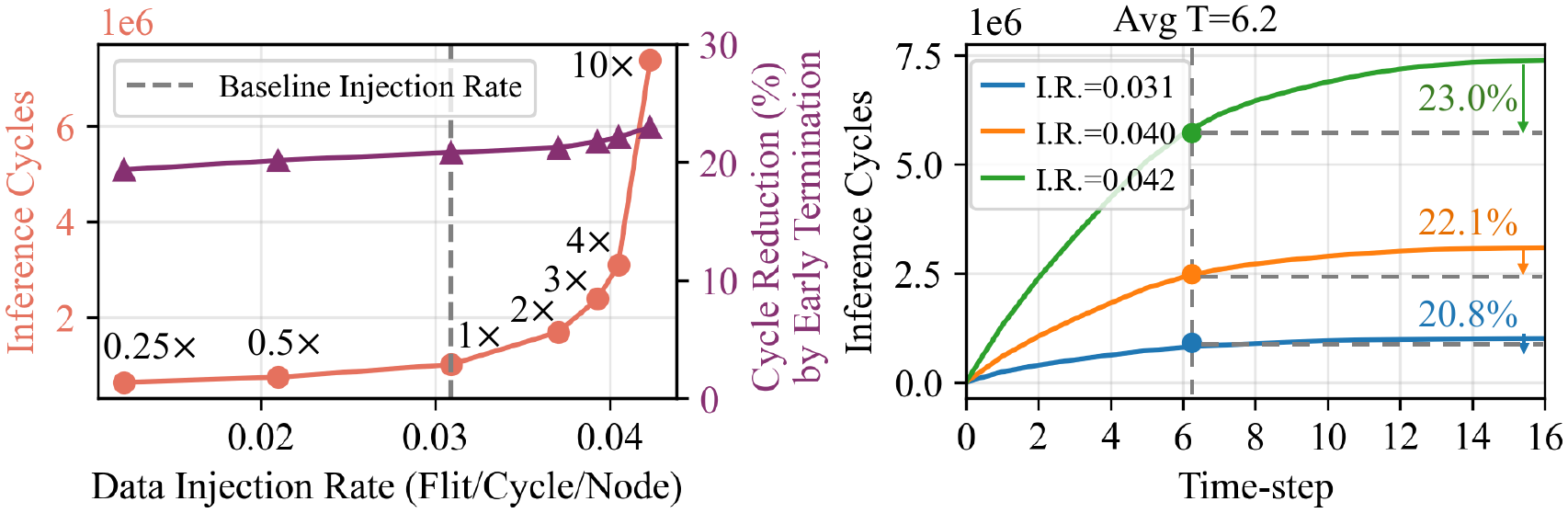}
    \caption{\hl{Total inference cycles together with the cycle reduction achieved by early termination (left) and average inference cycles per time-step (right) under different injection rates. 
    SNN is ResNet50, and the confidence threshold is 0.55. “1×” in the left figure denotes \#flits normalized to the baseline.}}
    \label{fig:Injection_rate}
\end{figure}


\subsection{Network Congestion Analysis}

\hl{To analyze on-chip network congestion, we vary the number of flits by adjusting the number of effective spikes packed into each flit. We then measure the inference cycles under different data injection rates, as shown in \cref{fig:Injection_rate}. All flits have a fixed size of 512 bits. The gray dotted line in \cref{fig:Injection_rate} (left) marks the baseline injection rate of 0.031 for ResNet50 in practical cases.
As shown in \cref{fig:Injection_rate} (left), when the injection rate exceeds 0.04, which is the 10$\times$ number of flits in the practical cases, the on-chip network becomes congested, and the inference cycles increase dramatically.
Nevertheless, the cycle reduction achieved by elastic inference remains stable and is always larger than 19\%.
This indicates that the network congestion does not affect the benefit of elastic inference.
\cref{fig:Injection_rate} (right) explains the reason.
Under different injection rates, the cycles of all time-steps increase proportionally, rather than being stalled at the first time-step.}

\subsection{Ablation Study}
This section explores the factors contributing to \sysname's high throughput and energy efficiency, including Gustavson-product, bundled AER, spine/token-wise pipeline, multi-path routing algorithm, and memory technique.

\begin{figure}[t]
    \centering
    \includegraphics[width=\linewidth]{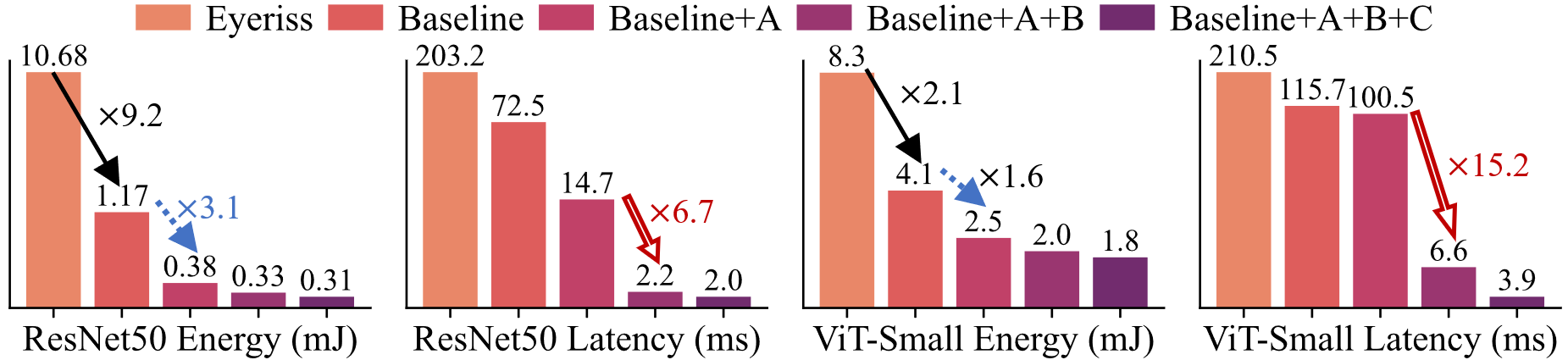}
    \vspace{-1.5em}
    \caption{Breakdown of techniques. A: Gustavson-product, B: spine/token-wise pipeline, and C: BAER. Baseline denotes \sysname without any optimizations and pipeline scheduling.}
    \vspace{-1em}
    \label{fig:ablution_study}
\end{figure}

\subsubsection{Technique Breakdown}
\label{sec:ablution_study}
\cref{fig:ablution_study} presents the energy and latency of various technique combinations in \sysname. Our baseline includes near-memory computing, addition-only computation, and large on-chip SRAM, but no architectural optimizations (\ie, inner-product, normal AER, and no pipeline scheduling). With these naive optimizations, our baseline outperforms Eyeriss \cite{7738524}, saving $9.2\times$ energy on ResNet50 and $2.1\times$ on ViT-Small (black arrows). Introducing Gustavson-product (optimization-A) cuts energy further by $3.1\times$ and $1.6\times$ (blue dotted arrows), and also reduces weight‐buffer access cycles, improving latency. Adding the spine/token pipeline (optimization-B) delivers dramatic speed‐ups of $6.7\times$ on ResNet50 and $15.2\times$ on ViT-Small (red hollow arrows), which also lowers leakage energy consumed by on-chip SRAMs. Finally, bundled AER (optimization-C) yields modest latency gains, as most cycles remain dominated by PE computation rather than NoC traffic.

\subsubsection{Effectiveness of Gustavson-Product}
\label{sec:exp_VSA_PE}
\begin{figure}[t]
    \centering
    \includegraphics[width=\linewidth]{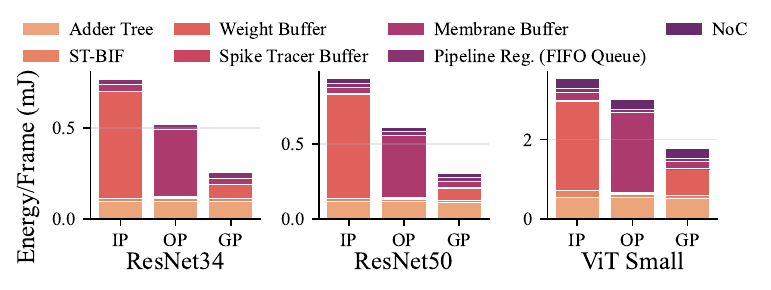}
    \vspace{-2.0em}
    \caption{Different Products benchmarked by ResNet34, ResNet50, ViT-Small on ImageNet. IP, OP, and GP denote Inner-Product, Outer-Product, and Gustavson-Product.}
    \vspace{-1.0em}
    \label{fig:product_dataflow}
\end{figure}
\begin{figure}[t]
\centering
    \includegraphics[width=\linewidth]{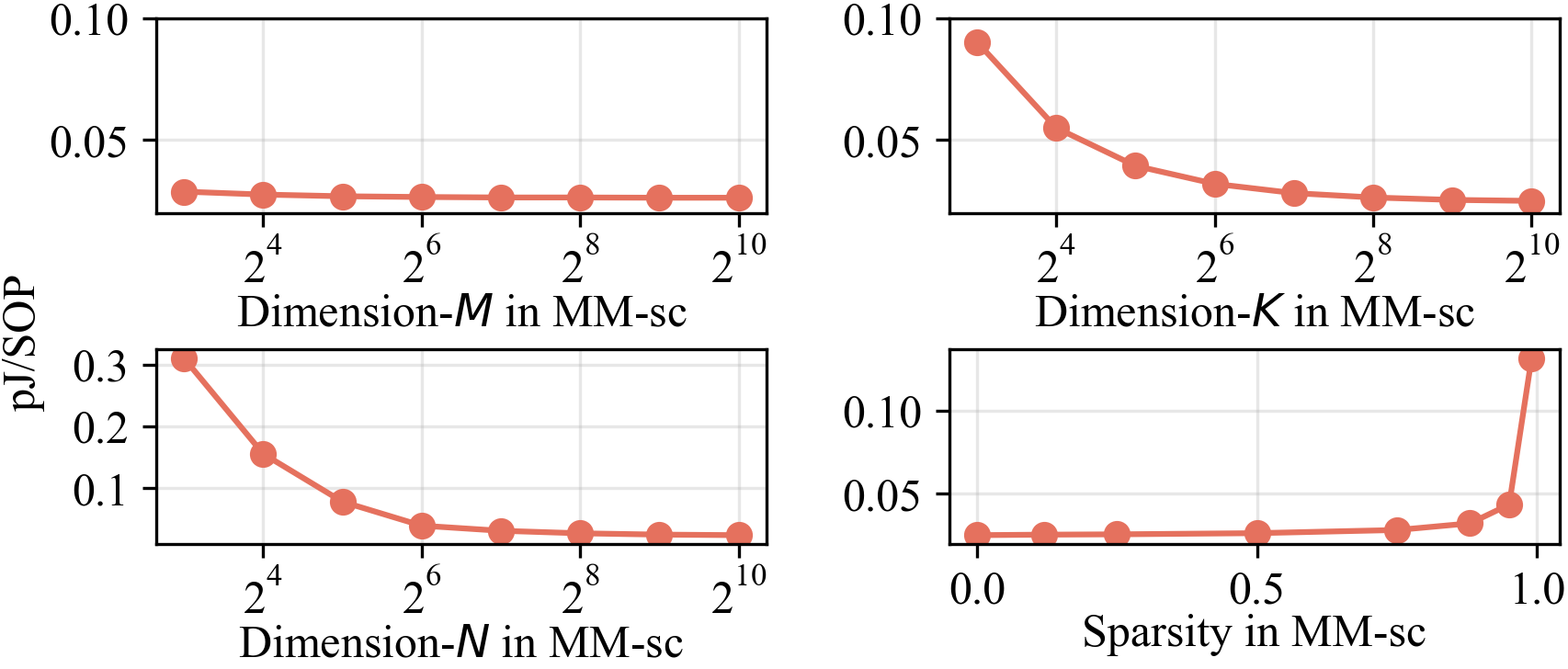}
    \caption{\hl{Energy (pJ/SOP) scaling with different matrix dimensions ($M \times K \times N$) and sparsity levels in MM-sc.}}
    \vspace{-1.5em}
    \label{fig:efficiency_GP}
\end{figure}
To highlight the benefits of the Gustavson-product, \cref{fig:product_dataflow} breaks down energy for three sparse, event-driven product algorithms: inner-product, outer-product, and Gustavson-product. All three incur similar adder‐tree costs. Inner-product suffers from high weight‐buffer energy (76.2\% on ResNet34) because it reads the full dense weight matrix to generate a membrane potential row. Outer-product minimizes weight‐buffer use (1.3\%) but repeatedly accesses the high‐precision membrane buffer, which dominates energy (70.3\%). Gustavson-product combines sparse weight reads with membrane‐stationary updates, cutting combined buffer energy to 43.1\%. Across ResNet34, ResNet50, and ViT-Small, this yields average savings of $2.7\times$ versus inner-product and $1.9\times$ versus outer-product. \hl{As shown in \cref{fig:efficiency_GP}, the energy efficiency of Gustavson-product is sensitive to the $K$ and $N$ dimensions, varying from 0.31 to 0.023 pJ/SOP for $K$ and from 0.09 to 0.025 pJ/SOP for $N$. This sensitivity arises because dimension $K$ determines the spike batching size; a small $K$ reduces adder utilization and increases the memory access overhead amortized per spike. Meanwhile, a small $N$ under-utilizes the SRAM bandwidth (64-bit by default), thereby degrading energy efficiency. For sparsity, since small sparsity increases hardware utilization, the energy cost (pJ/SOP) decreases. }

\begin{figure}[t]
    \centering
    \includegraphics[width=\linewidth]{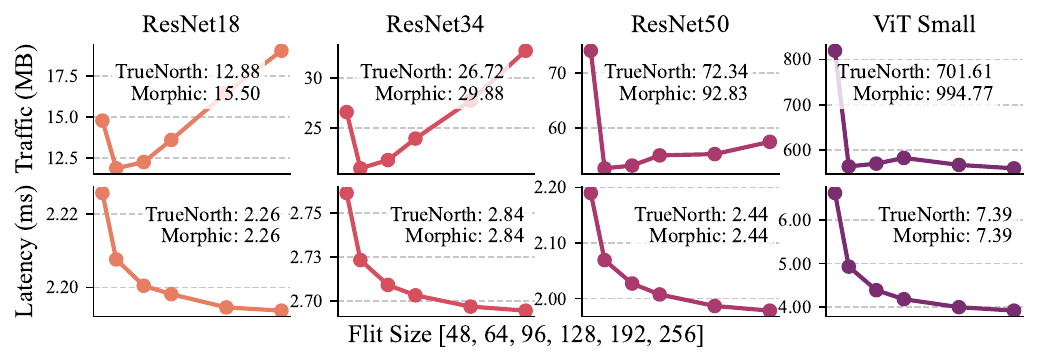}
    \vspace{-2.0em}
    \caption{NoC Traffic and Latency Across Various Flit Sizes.}
    \label{fig:flitSize}
    \vspace{-1.0em}
\end{figure}

\subsubsection{Effectiveness of Bundled AER}
\label{sec:exp_BAER}
Bundled AER stores the spine/token ID across spikes only once, cutting NoC traffic.
\cref{fig:flitSize} plots traffic and latency versus flit size, and compares \sysname to TrueNorth \cite{akopyan2015truenorth} and MorphIC \cite{morphic}.
At 64-bit flit size, bundled AER reduces traffic by an average 19.1\% versus TrueNorth and 36.7\% versus MorphIC.
TrueNorth has less traffic than MorphIC, since TrueNorth discards a 3-bit X-dimension hop count when the hop number is 0.
Additionally, in \cref{fig:flitSize}, we observe that: 1) Across models, traffic first falls then rises with flit size. 
Small flits (\eg 48 bits) split each spine/token into many flits, inflating traffic.
Very large flits (\eg 256 bits) under-utilize payload, also raising traffic. 
This phenomenon is lessened in models with more spikes per spine/token, such as ResNet50 and ViT-S, as the payload utilization in large flits is improved.
2) Latency steadily decreases as flit size grows, since fewer flits traverse the network. By contrast, TrueNorth and MorphIC send one spike per flit, generating many small flits and incurring higher latency.

\begin{figure}[t]
    \centering
    \includegraphics[width=\linewidth]{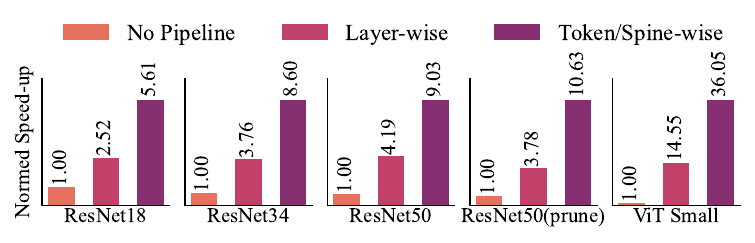}
    \vspace{-2.0em}
    \caption{Normalized speedup of \sysname using three different pipelines across various networks.}
    \label{fig:VSA_pipeline}
    \vspace{-1.5em}
\end{figure}

\begin{figure}[h]
    \centering
    \includegraphics[width=\linewidth]{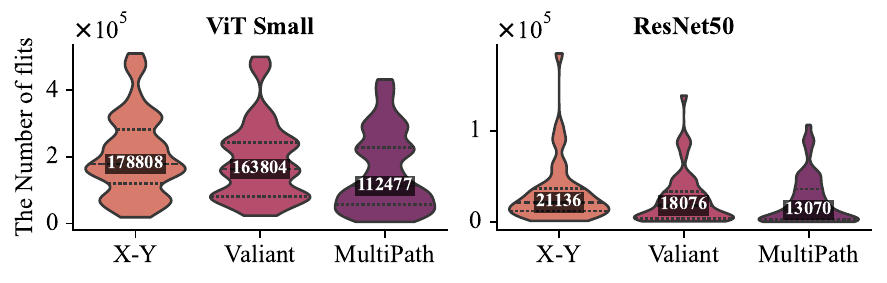}
    \vspace{-2em}
    \caption{Flit distribution across \sysname NoC links. Violin width shows flit count frequency; numbers mark \#flits medians, dotted lines indicate \#flits quartiles.}
    \vspace{-.5em}    
    \label{fig:NoC_transmit_Condition}
\end{figure}

\subsubsection{Effectiveness of Spine/Token-wise Pipeline}
\label{sec:exp_pipeline}
\cref{fig:VSA_pipeline} compares three pipelining strategies. 
The spine/token-wise pipeline achieves the lowest latency, yielding speedups of 2.2$\times$, 2.3$\times$, 2.8$\times$, and 2.5$\times$ over the no-pipeline baseline on ResNet18, ResNet34, ResNet50, and ViT-Small, respectively. 
By allowing each layer to start as soon as its spine or token is ready, hardware utilization is improved.
Note that the speedup in ViT-S (36.1$\times$) is more than ResNet18 (5.6$\times$), as ViT-S is deeper than ResNet18, leading to higher hardware utilization.

\subsubsection{Effectiveness of Multi-Path Routing Algorithm}
\label{sec:exp_Routing}
To evaluate our multi-path routing introduced in \cref{sec:mapping}, \cref{fig:NoC_transmit_Condition} plots the distribution of flits across NoC links. 
First, multi-path routing lowers required peak bandwidth by reducing the maximum flits per link, from 17.9 GB/s to 11.9 GB/s on ResNet50 and from 11.9 GB/s to 11.6 GB/s on ViT-S, compared to X-Y routing. Second, the distribution of flits with multi-path routing is more concentrated than Valiant's and X-Y routing algorithms, mitigating the communication imbalance.

\subsubsection{Effectiveness of Memory Technique}
\label{sec:exp_Memory}
\sysname to store all SNN parameters on-chip and performs frequent memory accesses during inference, making memory choice critical for area and energy efficiency. 
\cref{tab:memory_technology} compares SRAM and eDRAM implementations. SRAM delivers superior energy efficiency but consumes more area, while eDRAM reduces the area at the expense of higher energy cost. 
Designers can select the memory technology to make area–power trade-offs.

\begin{table}[h]
\vspace{-0.5em}
\caption{The TOPS/W and Allocated Area of ResNet18/34/50 with different Memory Techniques.}
\vspace{-0.5em}
\renewcommand\arraystretch{0.85}
\resizebox{\linewidth}{!}{
\begin{tabular}{@{}cccccc@{}}
\toprule
\textbf{} & \textbf{Metric} & \textbf{ResNet18} & \textbf{ResNet34} & \textbf{ResNet50} & \textbf{ViT Small} \\ \midrule
\textbf{SRAM} & \textbf{TOPS/W} & 29.96 & 27.12 & 25.55 & 5.10\\
\textbf{eDRAM} & \textbf{TOPS/W} & 5.97 & 5.32 & 3.64 & 2.61 \\
\textbf{SRAM} & \textbf{Area (mm\textsuperscript{2})} & 16.69 & 29.73 & 51.83 & 79.53 \\
\textbf{eDRAM} & \textbf{Area (mm\textsuperscript{2})} & 9.78 & 17.12 & 29.34 & 48.70 \\ \bottomrule
\end{tabular}}
\label{tab:memory_technology}
\vspace{-1.0em}
\end{table}

\subsubsection{Scaling Study}

\begin{figure}[t]
    \centering
    \includegraphics[width=\linewidth]{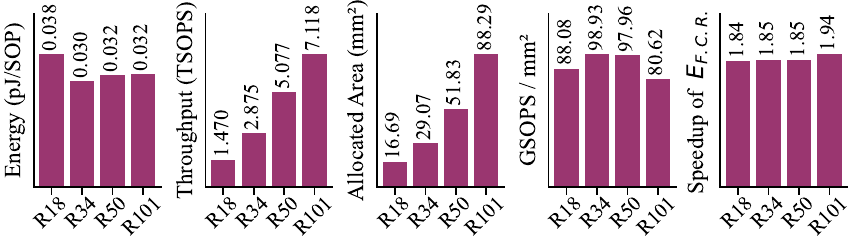}
    \vspace{-2em}
    \caption{\hl{Scaling study of \sysname~in ResNet18, ResNet34, ResNet50, and ResNet101,  including the energy (pJ/SOP), throughput (TSOPS), allocated area (mm$^2$), area efficiency (GOPS/mm$^2$), and speedup achieved by the expected latency of first-correct-response ($E_{\rm F.C.R}$). SOP is synaptic operation.
    }}
    \vspace{-1.5em}
    \label{fig:Scaling_Study}
\end{figure}

\begin{table}[t]
\caption{\hl{Scaling study about energy breakdown.} }
\vspace{-0.5em}
\label{tab:scaling_energy}
\renewcommand\arraystretch{0.8}
\resizebox{\linewidth}{!}{
\begin{tabular}{cccccc}
\toprule
Components & Detail & ResNet18 & ResNet34 & ResNet50 & \multicolumn{1}{l}{ResNet101} \\ \midrule
\multirow{4}{*}{Comput.} & Buffer & 52.58\% & 52.59\% & 50.31\% & 48.55\% \\
 & Adder & 34.68\% & 35.93\% & 35.70\% & 36.22\% \\
 & Neuron & 6.25\% & 4.77\% & 4.33\% & 4.56\% \\
 & Total & 93.50\% & 93.29\% & 90.34\% & 89.34\% \\ \midrule
\multirow{3}{*}{Communicat.} & NoC Traffic & 2.94\% & 4.09\% & 6.95\% & 8.02\% \\
 & Routing & 0.05\% & 0.03\% & 0.03\% & 0.03\% \\
 & Total & 2.99\% & 4.12\% & 6.98\% & 8.05\% \\ \midrule
\multirow{3}{*}{Scheduling} & Control+Scheduler & 1.21\% & 0.82\% & 1.12\% & 1.03\% \\
 & BAER & 2.29\% & 1.77\% & 1.56\% & 1.58\% \\
 & Total & 3.50\% & 2.59\% & 2.68\% & 2.61\% \\ \midrule
pJ/SOP & - & 0.038 & 0.030 & 0.032 & 0.032 \\ 
\bottomrule
\end{tabular}
}
\end{table}

\hl{\cref{fig:Scaling_Study} shows the scaling study of \sysname{} from ResNet18 to ResNet101. To eliminate the impact of sparsity, we use the number of synaptic operations (\#SOP) to calculate the throughput (TSOPS), energy efficiency (pJ/SOP), and area efficiency (GSOPS/mm$^2$). Overall, ELSA scales stably in energy efficiency (range from 0.030 to 0.038 pJ/SOP), area efficiency (range from 80.6 to 98.9 GOPS/mm$^2$), and speedup achieved by early correctness behavior (range from 1.84$\times$ to 1.94$\times$). Both throughput and area consumption increase with the model scaling. The reason is that a large model consumes more hardware resources, and the spine/token-wise pipeline further improves the throughput.}

\hl{
We also show the energy breakdown during the model size scaling in \cref{tab:scaling_energy}. 
Overall, the most energy consumption is concentrated on neural core computation (larger than 89.0\%).
With the model size scaling, the communication consumption increases from 2.99\% to 8.05\%.
The consumptions for supporting spine/token-level pipeline scheduling, including the BAER generator/decoder, PE control, and scheduler, are small ($<$ 4\%) and remain stable with the model scaling. 
}

\begin{table}[t]
\caption{\hl{SNN Executions Summary. “Time Advance” is the granularity at which components synchronously advance to the next time-step. ``S./T.",``Calcu.", ``Comm.", and ``Gran." are spine/token, calculation, communication, and granularity.}}
\vspace{-0.5em}
\label{tab:related_work_SNN_execution}
\renewcommand\arraystretch{0.9}
\resizebox{\linewidth}{!}{
\begin{tabular}{ccccc}
\hline
\multirow{2}{*}{Methods} & \multicolumn{2}{c}{Asynchronous Gran.} & \multirow{2}{*}{Time Advance} & \multirow{2}{*}{Schedule Gran.} \\ \cline{2-3}
 & Calcu. & Comm. &  &  \\ 
\midrule
Loihi\cite{davies2018loihi} & Spike & Spike & Chip-level & Network \\
SpiNNaker\cite{furber2014spinnaker} & Spike & Spike & Core-level & Layer \\
PAICORE\cite{PAICORE} & Spike & Spike & Core-level & Layer \\
\sysname (Ours) & S./T. & S./T. & PE-level & S./T. \\ 
\hline
\end{tabular}
}
\vspace{-1.0em}
\end{table}

\section{Related Work}
\label{sec:related_works}

In this section, we make a comprehensive discussion of existing digital, analog, and in-/near-memory accelerators to better position ELSA within a broad accelerator landscape.

\subsection{ANN Accelerators}
\textbf{Digital Designs}\cite{GAMMA,MatRaptor} also use Gustavson product to accelerate sparse-aware ANN by reducing memory access frequency, mitigating memory bottlenecks compared to inner-/outer-product methods.
Similarly, dataflow architectures such as Groq\cite{Groq} and Cerebras\cite{lie2023cerebras} employ near-SRAM designs, storing weights on-chip to minimize memory overhead.
Recent SRAM-based in-/near-memory design \cite{polymorpic,Nie2024VSPIM,MAICC} further exploit on-chip SRAM arrays for neural-network computation.
\sysname~builds upon these techniques, incorporating SNN-specific optimizations, as discussed in \cref{sec:snn_motivation}.

\textbf{Analog In-memory Designs}\cite{LLH-CIM,liu202033} use memristive crossbar arrays to accelerate GEMM by computing directly within memory. However, the multi-row accumulation in Gustavson product may suffer from non-idealities like IR drop\cite{fouda2020ir} and conductance discretization\cite{jain2020rxnn} in crossbar arrays, which impact accuracy. Thus, \sysname~adopts a fully digital design instead of an analog in-memory implementation.

\subsection{SNN Accelerators}

\textbf{Elastic SNN Accelerators}\cite{akopyan2015truenorth, davies2018loihi, darwin3, PAICORE, spinnaker2},\cite{hala_point_intel2024} follow TBT execution, featuring elastic inference with progressively emerged outputs. 
These accelerators generally adopt a multi-core design\cite{davies2018loihi} and exploit optimizations such as event-driven \cite{darwin3}, addition-only\cite{akopyan2015truenorth}, near-SRAM\cite{hala_point_intel2024} and dataflow architecture\cite{PAICORE}. 
However, their underlying coarse and layer-wise pipeline\cite{PAICORE} fundamentally limits the exploitation of the early-response by elastic inference, as illustrated in \cref{fig:pipeline_comparison}.
\cref{tab:related_work_SNN_execution} summarizes the execution differences between ELSA and prior elastic accelerators.
\sysname~adopts spine/token-level granularity for both computation and communication, unlike the spike-level granularity used in Loihi\cite{davies2018loihi}, SpiNNaker\cite{furber2014spinnaker}, and PAICORE\cite{PAICORE}. 
Moreover, \sysname~advances the time-step at the PE-level, finer than other related works, allowing each neural core to manage the spines/tokens independently.

\textbf{Non-elastic SNN Accelerators}~\cite{SASAP, narayanan2020spinalflow, prosperity, phi, C-DNN, hwang2026gustavsnn} follow LBL execution to exploit time-step parallelism and avoid costly membrane SRAM storage, thereby improving throughput and energy efficiency. Specifically, SASAP\cite{SASAP} scales up compute units to process spikes across all time-steps in parallel, while membrane states are immediately discarded after use\cite{hwang2026gustavsnn}.
However, such execution is inherently incompatible with early-response since outputs are synchronously generated before proceeding to the next layer.

\section{Conclusion}

This paper presents \sysname, a near-SRAM dataflow architecture specifically designed to exploit elastic inference.
By enabling early responses, \sysname is better suited for real-time applications such as autonomous driving compared to prior SNN accelerators.
Experimentally, \sysname demonstrates that SNNs can outperform QANNs while maintaining comparable accuracy, reinforcing the potential of SNNs for future high-performance, low-power applications.

\appendix
\section{Artifact Appendix}
\subsection{Abstract}
Our artificial evaluation has two major parts: the evaluation of the SNN model accuracy and the performance of the ELSA. 

We evaluate our results using SNN models on standard image classification tasks. The evaluation encompasses five representative models: VGG16, ResNet18, ResNet34, ResNet50, and ViT-Small, and three widely-used datasets: CIFAR-10, CIFAR-100, and ImageNet. To facilitate reproducibility, we provide validation scripts and pre-trained checkpoints for all models, allowing rapid accuracy verification.

For assessing ELSA performance, we adopt a two-path evaluation strategy: a slow path and a fast path. In the slow path, the simulator generates energy and latency tracer files, from which power, performance, and area (PPA) metrics are computed. This process requires approximately 8 hours. In the fast path, we provide the pre-generated tracer files, allowing direct computation of PPA metrics within one minute.

All experiments are conducted on an Ubuntu server equipped with eight NVIDIA RTX 4090 GPUs, ensuring consistent and high-performance evaluation across all models.

\subsection{Artifact check-list (meta-information)}

{\small
\begin{itemize}
  \item {\bf Compilation:} GCC: 11.4.0
  \item {\bf Model:} VGG-16, ResNet18, ResNet34, ResNet50, and ViT Small.
  \item {\bf Data set: }ImageNet, CIFAR10, CIFAR100.
  \item {\bf Run-time environment: } Ubuntu 22.04.3 LTS, CUDA 12.2, and PyTorch 2.4.1.
  \item {\bf Hardware: } A server with an AMD EPYC 9334 32-Core Processor and eight NVIDIA 4090 GPUs.
  \item {\bf Output: } SNN accuracy, ELSA energy, performance, and area.
  \item {\bf How much disk space is required (approximately)?: } 20GB.
  \item {\bf How much time is needed to prepare the workflow (approximately)?: } It takes about 30 minutes to prepare the environment.
  \item {\bf How much time is needed to complete experiments (approximately)?: } Obtaining ELSA PPA metrics requires approximately 8 hours, and evaluating the SNN model accuracy also takes around 8 hours. Using the pre-generated tracer files, the fast evaluation of ELSA PPA metrics can be completed in under one minute.
  \item {\bf Publicly available: } Our framework is publicly available on GitHub \url{https://github.com/Intelligent-Computing-Research-Group/ELSA#}.
  \item {\bf Data licenses:} The datasets are publicly available through their original licensing terms.
  \item {\bf Archived: } \url{https://zenodo.org/records/19449728}.
\end{itemize}
}

\subsection{Description}

\subsubsection{How to access}

We archive the source code at \url{https://zenodo.org/records/19449728}. We recommend you access the provided anonymous GitHub repository: \url{https://github.com/Intelligent-Computing-Research-Group/ELSA#} for the latest version.

\subsubsection{Hardware dependencies}
We evaluate the SNN models with two types of server configuration: \ding{172} ELSA simulator: a server equipped with an AMD EPYC 9334 32-Core Processor. \ding{173} ELSA algorithm evaluator: a server equipped with eight NVIDIA 4090 GPUs.
\subsubsection{Software dependencies}
The experiments rely on the following software components.
\begin{itemize}
    \item Ubuntu 22.04.3 LTS
    \item Python 3.10
    \item PyTorch 2.4.1
    \item Anaconda 24.5.0
    \item GCC 11.4.0
    \item CUDA 12.2
\end{itemize}
\subsubsection{Data sets and models}
The evaluated image classification models with the ImageNet dataset~\cite{deng2009imagenet}, CIFAR10 dataset, and CIDAR100 dataset.
We evaluate the SNN models including VGG-16~\cite{Simonyan2014VeryDC}, ResNet-18~\cite{resnet},ResNet-34~\cite{resnet}, ResNet-50~\cite{resnet}, and ViT (vision transformer)~\cite{dosovitskiy2020image}. 

\subsection{Installation}

We have  well-documented README files to detail the installation instructions for each experiment at 
\url{https://github.com/Intelligent-Computing-Research-Group/ELSA#}.

\subsection{Evaluation and expected results}

Our experiments have two major parts: the evaluation of SNN model accuracy and the performance of the ELSA accelerator.

\begin{itemize}
    \item The directory \texttt{ELSA\_Algorithm} contains the ELSA framework based on PyTorch for the SNN model accuracy and elastic inference evaluation.
    \item The directory \texttt{ELSA\_Simulator} contains the performance and energy evaluation of the ELSA simulator.
\end{itemize}

To evaluate the experiments, you can utilize the scripts in each directory according to the README files.
We also release all expected results in the README files for Figure~\ref{fig:QANN_comparsion}, and Figure~\ref{SNN_benchmarks}.

\subsection{Methodology}

Submission, reviewing, and badging methodology:

\begin{itemize}
  \item Submission instructions: \url{https://iscaconf.org/isca2026/submit/artifactevaluation.php}
  \item Reviewing process: \url{https://github.com/ctuning/artifact-evaluation/blob/master/docs/reviewing.md}
  \item Artifact Review \& Badging policy: \url{https://www.acm.org/publications/policies/artifact-review-and-badging-current}
\end{itemize}

\newpage

\bibliographystyle{IEEEtran}
\bibliography{refs}

@article{zhou2022spikformer,
  title={Spikformer: When spiking neural network meets transformer},
  author={Zhou, Zhaokun and Zhu, Yuesheng and He, Chao and Wang, Yaowei and Yan, Shuicheng and Tian, Yonghong and Yuan, Li},
  journal={arXiv preprint arXiv:2209.15425},
  year={2022}
}

@article{hu2023fast,
  title={Fast-snn: Fast spiking neural network by converting quantized ann},
  author={Hu, Yangfan and Zheng, Qian and Jiang, Xudong and Pan, Gang},
  journal={IEEE Transactions on Pattern Analysis and Machine Intelligence},
  year={2023},
  publisher={IEEE}
}

@inproceedings{
spikeziptf2024,
title={SpikeZIP-TF: Conversion is All You Need for Transformer-based SNN},
author={You, Kang and Xu, Zekai and Nie, Chen and Deng, Zhijie and Wang, Xiang and Guo, Qinghai and He, Zhezhi},
booktitle={Forty-first International Conference on Machine Learning (ICML)},
year={2024}
}

@article{davies2018loihi,
  title={Loihi: A neuromorphic manycore processor with on-chip learning},
  author={Davies, Mike and Srinivasa, Narayan and Lin, Tsung-Han and Chinya, Gautham and Cao, Yongqiang and Choday, Sri Harsha and Dimou, Georgios and Joshi, Prasad and Imam, Nabil and Jain, Shweta},
  journal={Ieee Micro},
  volume={38},
  number={1},
  pages={82--99},
  year={2018},
  publisher={IEEE}
}

@article{akopyan2015truenorth,
  title={Truenorth: Design and tool flow of a 65 mw 1 million neuron programmable neurosynaptic chip},
  author={Akopyan, Filipp and Sawada, Jun and Cassidy, Andrew and Alvarez-Icaza, Rodrigo and Arthur, John and Merolla, Paul and Imam, Nabil and Nakamura, Yutaka and Datta, Pallab and Nam, Gi-Joon},
  journal={IEEE transactions on computer-aided design of integrated circuits and systems},
  volume={34},
  number={10},
  pages={1537--1557},
  year={2015},
  publisher={IEEE}
}

@article{bu2023optimal,
  title={Optimal ANN-SNN conversion for high-accuracy and ultra-low-latency spiking neural networks},
  author={Bu, Tong and Fang, Wei and Ding, Jianhao and Dai, PengLin and Yu, Zhaofei and Huang, Tiejun},
  journal={arXiv preprint arXiv:2303.04347},
  year={2023}
}

@inproceedings{lee2020reconfigurable,
  title={Reconfigurable dataflow optimization for spatiotemporal spiking neural computation on systolic array accelerators},
  author={Lee, Jeong-Jun and Li, Peng},
  booktitle={2020 IEEE 38th International Conference on Computer Design (ICCD)},
  pages={57--64},
  year={2020},
  organization={IEEE}
}

@inproceedings{liu2022sato,
  title={Sato: spiking neural network acceleration via temporal-oriented dataflow and architecture},
  author={Liu, Fangxin and Zhao, Wenbo and Wang, Zongwu and Chen, Yongbiao and Yang, Tao and He, Zhezhi and Yang, Xiaokang and Jiang, Li},
  booktitle={Proceedings of the 59th ACM/IEEE Design Automation Conference},
  pages={1105--1110},
  year={2022}
}

@inproceedings{narayanan2020spinalflow,
  title={SpinalFlow: An architecture and dataflow tailored for spiking neural networks},
  author={Narayanan, Surya and Taht, Karl and Balasubramonian, Rajeev and Giacomin, Edouard and Gaillardon, Pierre-Emmanuel},
  booktitle={2020 ACM/IEEE 47th Annual International Symposium on Computer Architecture (ISCA)},
  pages={349--362},
  year={2020},
  organization={IEEE}
}

@inproceedings{deng2009imagenet,
  title={Imagenet: A large-scale hierarchical image database},
  author={Deng, Jia and Dong, Wei and Socher, Richard and Li, Li-Jia and Li, Kai and Fei-Fei, Li},
  booktitle={2009 IEEE conference on computer vision and pattern recognition},
  pages={248--255},
  year={2009},
  organization={Ieee}
}

@TECHREPORT{alex2009cifar,
            author={Alex Krizhevsky},
            title={Learning multiple layers of features from tiny images},
            institution={},
            year={2009}
}

@article{li2017cifar10dvs,
  title={CIFAR10-DVS: An Event-Stream Dataset for Object Classification},
  author={Li, Hongmin  and Liu, Hanchao  and Ji, Xiangyang  and Li, Guoqi  and Shi, Luping},
  journal={Frontiers in Neuroscience},
  volume={11},
  year={2017},
}

@article{roy2019towards,
  title={Towards spike-based machine intelligence with neuromorphic computing},
  author={Roy, Kaushik and Jaiswal, Akhilesh and Panda, Priyadarshini},
  journal={Nature},
  volume={575},
  number={7784},
  pages={607--617},
  year={2019},
  publisher={Nature Publishing Group UK London}
}

@inproceedings{mao2024stellar,
  title={Stellar: Energy-Efficient and Low-Latency SNN Algorithm and Hardware Co-Design with Spatiotemporal Computation},
  author={Mao, Ruixin and Tang, Lin and Yuan, Xingyu and Liu, Ye and Zhou, Jun},
  booktitle={2024 IEEE International Symposium on High-Performance Computer Architecture (HPCA)},
  pages={172--185},
  year={2024},
  organization={IEEE}
}

@inproceedings{lee2022parallel,
  title={Parallel time batching: Systolic-array acceleration of sparse spiking neural computation},
  author={Lee, Jeong-Jun and Zhang, Wenrui and Li, Peng},
  booktitle={2022 IEEE International Symposium on High-Performance Computer Architecture (HPCA)},
  pages={317--330},
  year={2022},
  organization={IEEE}
}

@ARTICLE{7738524,
  author={Chen, Yu-Hsin and Krishna, Tushar and Emer, Joel S. and Sze, Vivienne},
  journal={IEEE Journal of Solid-State Circuits}, 
  title={Eyeriss: An Energy-Efficient Reconfigurable Accelerator for Deep Convolutional Neural Networks}, 
  year={2017},
  volume={52},
  number={1},
  pages={127-138},
  doi={10.1109/JSSC.2016.2616357}}

@INPROCEEDINGS{resnet,
  author={He, Kaiming and Zhang, Xiangyu and Ren, Shaoqing and Sun, Jian},
  booktitle={2016 IEEE Conference on Computer Vision and Pattern Recognition (CVPR)}, 
  title={Deep Residual Learning for Image Recognition}, 
  year={2016},
  volume={},
  number={},
  pages={770-778},
  doi={10.1109/CVPR.2016.90}}

@inproceedings{visionTransformer,
  author       = {Alexey Dosovitskiy and
                  Lucas Beyer and
                  Alexander Kolesnikov and
                  Dirk Weissenborn and
                  Xiaohua Zhai and
                  Thomas Unterthiner and
                  Mostafa Dehghani and
                  Matthias Minderer and
                  Georg Heigold and
                  Sylvain Gelly and
                  Jakob Uszkoreit and
                  Neil Houlsby},
  title        = {An Image is Worth 16x16 Words: Transformers for Image Recognition
                  at Scale},
  booktitle    = {9th International Conference on Learning Representations, {ICLR} 2021,
                  Virtual Event, Austria, May 3-7, 2021},
  publisher    = {OpenReview.net},
  year         = {2021},
  url          = {https://openreview.net/forum?id=YicbFdNTTy},
  bibsource    = {dblp computer science bibliography, https://dblp.org}
}

@article{Simonyan2014VeryDC,
  title={Very Deep Convolutional Networks for Large-Scale Image Recognition},
  author={Karen Simonyan and Andrew Zisserman},
  journal={CoRR},
  year={2014},
  volume={abs/1409.1556},
  url={https://api.semanticscholar.org/CorpusID:14124313}
}

@ARTICLE{C-DNN,
  author={Kim, Sangyeob and Kim, Soyeon and Hong, Seongyon and Kim, Sangjin and Han, Donghyeon and Choi, Jiwon and Yoo, Hoi-Jun},
  journal={IEEE Journal of Solid-State Circuits}, 
  title={C-DNN: An Energy-Efficient Complementary Deep-Neural-Network Processor With Heterogeneous CNN/SNN Core Architecture}, 
  year={2024},
  volume={59},
  number={1},
  pages={157-172},
  doi={10.1109/JSSC.2023.3330483}}

@INPROCEEDINGS{S-CONV,
  author={Sun, Wenyu and Feng, Xiaoyu and Tang, Chen and Fan, Shupei and Yang, Yixiong and Yue, Jinshan and Yang, Huazhong and Liu, Yongpan},
  booktitle={2023 IEEE International Solid-State Circuits Conference (ISSCC)}, 
  title={A 28nm 2D/3D Unified Sparse Convolution Accelerator with Block-Wise Neighbor Searcher for Large-Scaled Voxel-Based Point Cloud Network}, 
  year={2023},
  volume={},
  number={},
  pages={328-330},
  doi={10.1109/ISSCC42615.2023.10067644}}

@INPROCEEDINGS{ANT,
  author={Guo, Cong and Zhang, Chen and Leng, Jingwen and Liu, Zihan and Yang, Fan and Liu, Yunxin and Guo, Minyi and Zhu, Yuhao},
  booktitle={2022 55th IEEE/ACM International Symposium on Microarchitecture (MICRO)}, 
  title={ANT: Exploiting Adaptive Numerical Data Type for Low-bit Deep Neural Network Quantization}, 
  year={2022},
  volume={},
  number={},
  pages={1414-1433},
  doi={10.1109/MICRO56248.2022.00095}}

@INPROCEEDINGS{BitFusion,
  author={Sharma, Hardik and Park, Jongse and Suda, Naveen and Lai, Liangzhen and Chau, Benson and Kim, Joon Kyung and Chandra, Vikas and Esmaeilzadeh, Hadi},
  booktitle={2018 ACM/IEEE 45th Annual International Symposium on Computer Architecture (ISCA)}, 
  title={Bit Fusion: Bit-Level Dynamically Composable Architecture for Accelerating Deep Neural Network}, 
  year={2018},
  volume={},
  number={},
  pages={764-775},
  doi={10.1109/ISCA.2018.00069}}

@INPROCEEDINGS{AICAS2024,
  author={Lin, Cheng-Chen and Lu, Wei and Huang, Po-Tsang and Chen, Hung-Ming},
  booktitle={2024 IEEE 6th International Conference on AI Circuits and Systems (AICAS)}, 
  title={A 28nm 343.5fps/W Vision Transformer Accelerator with Integer-Only Quantized Attention Block}, 
  year={2024},
  volume={},
  number={},
  pages={80-84},
  doi={10.1109/AICAS59952.2024.10595969}}

@INPROCEEDINGS{ViTALiTy,
  author={Dass, Jyotikrishna and Wu, Shang and Shi, Huihong and Li, Chaojian and Ye, Zhifan and Wang, Zhongfeng and Lin, Yingyan},
  booktitle={2023 IEEE International Symposium on High-Performance Computer Architecture (HPCA)}, 
  title={ViTALiTy: Unifying Low-rank and Sparse Approximation for Vision Transformer Acceleration with a Linear Taylor Attention}, 
  year={2023},
  volume={},
  number={},
  pages={415-428},
  doi={10.1109/HPCA56546.2023.10071081}}

@inproceedings{Sanger,
author = {Lu, Liqiang and Jin, Yicheng and Bi, Hangrui and Luo, Zizhang and Li, Peng and Wang, Tao and Liang, Yun},
title = {Sanger: A Co-Design Framework for Enabling Sparse Attention using Reconfigurable Architecture},
year = {2021},
isbn = {9781450385572},
publisher = {Association for Computing Machinery},
address = {New York, NY, USA},
url = {https://doi.org/10.1145/3466752.3480125},
doi = {10.1145/3466752.3480125},
booktitle = {MICRO-54: 54th Annual IEEE/ACM International Symposium on Microarchitecture},
pages = {977–991},
numpages = {15},
location = {Virtual Event, Greece},
series = {MICRO '21}
}

@article{dosovitskiy2020image,
  title={An image is worth 16x16 words: Transformers for image recognition at scale},
  author={Dosovitskiy, Alexey},
  journal={arXiv preprint arXiv:2010.11929},
  year={2020}
}

@inproceedings{2D_MESH,
  title={Simulation and analysis of network on chip architectures: ring, spidergon and 2D mesh},
  author={Bononi, Luciano and Concer, Nicola},
  booktitle={Proceedings of the Design Automation \& Test in Europe Conference},
  volume={2},
  pages={6--pp},
  year={2006},
  organization={IEEE}
}

@inproceedings{marchisio2023swifttron,
  title={SwiftTron: An efficient hardware accelerator for quantized transformers},
  author={Marchisio, Alberto and Dura, Davide and Capra, Maurizio and Martina, Maurizio and Masera, Guido and Shafique, Muhammad},
  booktitle={2023 International Joint Conference on Neural Networks (IJCNN)},
  pages={1--9},
  year={2023},
  organization={IEEE}
}

@article{li2020dramsim3,
  title={DRAMsim3: A cycle-accurate, thermal-capable DRAM simulator},
  author={Li, Shang and Yang, Zhiyuan and Reddy, Dhiraj and Srivastava, Ankur and Jacob, Bruce},
  journal={IEEE Computer Architecture Letters},
  volume={19},
  number={2},
  pages={106--109},
  year={2020},
  publisher={IEEE}
}

@inproceedings{o2014highlights,
  title={Highlights of the high-bandwidth memory (hbm) standard},
  author={O’Connor, Mike},
  booktitle={Memory forum workshop},
  volume={3},
  year={2014}
}

@article{mittal2014survey,
  title={A survey of architectural approaches for managing embedded DRAM and non-volatile on-chip caches},
  author={Mittal, Sparsh and Vetter, Jeffrey S and Li, Dong},
  journal={IEEE Transactions on Parallel and Distributed Systems},
  volume={26},
  number={6},
  pages={1524--1537},
  year={2014},
  publisher={IEEE}
}

@article{matick2005logic,
  title={Logic-based eDRAM: Origins and rationale for use},
  author={Matick, Richard E and Schuster, Stanley E},
  journal={IBM Journal of Research and Development},
  volume={49},
  number={1},
  pages={145--165},
  year={2005},
  publisher={IBM}
}

@inproceedings{han2020rmp,
  title={Rmp-snn: Residual membrane potential neuron for enabling deeper high-accuracy and low-latency spiking neural network},
  author={Han, Bing and Srinivasan, Gopalakrishnan and Roy, Kaushik},
  booktitle={Proceedings of the IEEE/CVF conference on computer vision and pattern recognition},
  pages={13558--13567},
  year={2020}
}

@article{qian20223d,
  title={3D object detection for autonomous driving: A survey},
  author={Jiageng Mao and Shaoshuai Shi and Xiaogang Wang and Hongsheng Li},
  journal={Pattern Recognition},
  volume={130},
  pages={108796},
  year={2022},
  publisher={Elsevier}
}

@inproceedings{yin2024loas,
  title={LoAS: Fully Temporal-Parallel Dataflow for Dual-Sparse Spiking Neural Networks},
  author={Yin, Ruokai and Kim, Youngeun and Wu, Di and Panda, Priyadarshini},
  booktitle={2024 57th IEEE/ACM International Symposium on Microarchitecture (MICRO)},
  pages={1107--1121},
  year={2024},
  organization={IEEE}
}

@ARTICLE{PAICORE,
  author={Zhong, Yi and Kuang, Yisong and Liu, Kefei and Wang, Zilin and Feng, Shuo and Chen, Guang and Yang, Youming and Cui, Xiuping and Wang, Qiankun and Cao, Jian and Jia, Song and Liang, Yun and Sun, Guangyu and Cui, Xiaoxin and Huang, Ru and Wang, Yuan},
  journal={IEEE Journal of Solid-State Circuits}, 
  title={PAICORE: A 1.9-Million-Neuron 5.181-TSOPS/W Digital Neuromorphic Processor With Unified SNN-ANN and On-Chip Learning Paradigm}, 
  year={2025},
  volume={60},
  number={2},
  pages={651-671},
  doi={10.1109/JSSC.2024.3426319}}

@inproceedings{Mapping_Zheda,
author = {Jin, Ouwen and Xing, Qinghui and Li, Ying and Deng, Shuiguang and He, Shuibing and Pan, Gang},
title = {Mapping Very Large Scale Spiking Neuron Network to Neuromorphic Hardware},
year = {2023},
isbn = {9781450399180},
publisher = {Association for Computing Machinery},
address = {New York, NY, USA},
url = {https://doi.org/10.1145/3582016.3582038},
doi = {10.1145/3582016.3582038},
booktitle = {Proceedings of the 28th ACM International Conference on Architectural Support for Programming Languages and Operating Systems, Volume 3},
pages = {419–432},
numpages = {14},
location = {Vancouver, BC, Canada},
series = {ASPLOS 2023}
}

@Article{modified_hilert_curve,
AUTHOR = {Rong, Yibiao and Zhang, Xia and Lin, Jianyu},
TITLE = {Modified Hilbert Curve for Rectangles and Cuboids and Its Application in Entropy Coding for Image and Video Compression},
JOURNAL = {Entropy},
VOLUME = {23},
YEAR = {2021},
NUMBER = {7},
ARTICLE-NUMBER = {836},
URL = {https://www.mdpi.com/1099-4300/23/7/836},
PubMedID = {34210074},
ISSN = {1099-4300},
DOI = {10.3390/e23070836}
}

@inproceedings{GAMMA,
author = {Zhang, Guowei and Attaluri, Nithya and Emer, Joel S. and Sanchez, Daniel},
title = {Gamma: leveraging Gustavson’s algorithm to accelerate sparse matrix multiplication},
year = {2021},
isbn = {9781450383172},
publisher = {Association for Computing Machinery},
address = {New York, NY, USA},
url = {https://doi.org/10.1145/3445814.3446702},
doi = {10.1145/3445814.3446702},
booktitle = {Proceedings of the 26th ACM International Conference on Architectural Support for Programming Languages and Operating Systems},
pages = {687–701},
numpages = {15},
location = {Virtual, USA},
series = {ASPLOS '21}
}

@INPROCEEDINGS{MatRaptor,
  author={Srivastava, Nitish and Jin, Hanchen and Liu, Jie and Albonesi, David and Zhang, Zhiru},
  booktitle={2020 53rd Annual IEEE/ACM International Symposium on Microarchitecture (MICRO)}, 
  title={MatRaptor: A Sparse-Sparse Matrix Multiplication Accelerator Based on Row-Wise Product}, 
  year={2020},
  volume={},
  number={},
  pages={766-780},
  doi={10.1109/MICRO50266.2020.00068}}

@misc{darwin3,
      title={Darwin3: A large-scale neuromorphic chip with a Novel ISA and On-Chip Learning}, 
      author={De Ma and Xiaofei Jin and Shichun Sun and Yitao Li and Xundong Wu and Youneng Hu and Fangchao Yang and Huajin Tang and Xiaolei Zhu and Peng Lin and Gang Pan},
      year={2023},
      eprint={2312.17582},
      archivePrefix={arXiv},
      primaryClass={cs.NE},
      url={https://arxiv.org/abs/2312.17582}, 
}

@INPROCEEDINGS{DSENT,
  author={Sun, Chen and Chen, Chia-Hsin Owen and Kurian, George and Wei, Lan and Miller, Jason and Agarwal, Anant and Peh, Li-Shiuan and Stojanovic, Vladimir},
  booktitle={2012 IEEE/ACM Sixth International Symposium on Networks-on-Chip}, 
  title={DSENT - A Tool Connecting Emerging Photonics with Electronics for Opto-Electronic Networks-on-Chip Modeling}, 
  year={2012},
  volume={},
  number={},
  pages={201-210},
  doi={10.1109/NOCS.2012.31}}

@INPROCEEDINGS{LLH-CIM,
  author={Guo, An and Chen, Xi and Dong, Fangyuan and Chen, Jinwu and Yuan, Zhihang and Hu, Xing and Zhang, Yuanpeng and Zhang, Jingmin and Tang, Yuchen and Zhang, Zhican and Chen, Gang and Yang, Dawei and Zhang, Zhaoyang and Ren, Lizheng and Xiong, Tianzhu and Wang, Bo and Liu, Bo and Shan, Weiwei and Liu, Xinning and Cai, Hao and Sun, Guangyu and Yang, Jun and Si, Xin},
  booktitle={2024 IEEE International Solid-State Circuits Conference (ISSCC)}, 
  title={34.3 A 22nm 64kb Lightning-Like Hybrid Computing-in-Memory Macro with a Compressed Adder Tree and Analog-Storage Quantizers for Transformer and CNNs}, 
  year={2024},
  volume={67},
  number={},
  pages={570-572},
  doi={10.1109/ISSCC49657.2024.10454278}}

@INPROCEEDINGS{AEC-CIM,
  author={Su, Jian-Wei and Chou, Yen-Chi and Liu, Ruhui and Liu, Ta-Wei and Lu, Pei-Jung and Wu, Ping-Chun and Chung, Yen-Lin and Hung, Li-Yang and Ren, Jin-Sheng and Pan, Tianlong and Li, Sih-Han and Chang, Shih-Chieh and Sheu, Shyh-Shyuan and Lo, Wei-Chung and Wu, Chih-I and Si, Xin and Lo, Chung-Chuan and Liu, Ren-Shuo and Hsieh, Chih-Cheng and Tang, Kea-Tiong and Chang, Meng-Fan},
  booktitle={2021 IEEE International Solid-State Circuits Conference (ISSCC)}, 
  title={16.3 A 28nm 384kb 6T-SRAM Computation-in-Memory Macro with 8b Precision for AI Edge Chips}, 
  year={2021},
  volume={64},
  number={},
  pages={250-252},
  doi={10.1109/ISSCC42613.2021.9365984}}

@INPROCEEDINGS{SASAP,
  author={Fang, Chaoming and Shen, Ziyang and Zhao, Shiqi and Wang, Chuanqing and Tian, Fengshi and Yang, Jie and Sawan, Mohamad},
  booktitle={2024 IEEE Custom Integrated Circuits Conference (CICC)}, 
  title={A 0.078 pJ/SOP Unstructured Sparsity-Aware Spiking Attention/Convolution Processor with 3D Compute Array}, 
  year={2024},
  volume={},
  number={},
  pages={1-2},
  doi={10.1109/CICC60959.2024.10529019}}

@ARTICLE{morphic,
  author={Frenkel, Charlotte and Legat, Jean-Didier and Bol, David},
  journal={IEEE Transactions on Biomedical Circuits and Systems}, 
  title={MorphIC: A 65-nm 738k-Synapse/mm$^2$ Quad-Core Binary-Weight Digital Neuromorphic Processor With Stochastic Spike-Driven Online Learning}, 
  year={2019},
  volume={13},
  number={5},
  pages={999-1010},
  doi={10.1109/TBCAS.2019.2928793}}

@online{Jetson_AGX_Orin,
author = {Nvidia},
title = {NVIDIA Jetson AGX Orin 64 GB},
url = {https://www.techpowerup.com/gpu-specs/jetson-agx-orin-64-gb.c4085},
note = {2021, Nov 09}
}

@online{Nvidia_A100,
author = {NVIDIA},
title = {NVIDIA A100},
url = {https://www.nvidia.cn/content/dam/en-zz/Solutions/Data-Center/a100/pdf/ampere-a100-datasheet-a4-nvidia-1293124-r10-web-zhCN.pdf},
note = {2020, May 04}
}

@inproceedings{TPUV4,
author = {Jouppi, Norm and Kurian, George and Li, Sheng and Ma, Peter and Nagarajan, Rahul and Nai, Lifeng and Patil, Nishant and Subramanian, Suvinay and Swing, Andy and Towles, Brian and Young, Clifford and Zhou, Xiang and Zhou, Zongwei and Patterson, David A},
title = {TPU v4: An Optically Reconfigurable Supercomputer for Machine Learning with Hardware Support for Embeddings},
year = {2023},
isbn = {9798400700958},
publisher = {Association for Computing Machinery},
address = {New York, NY, USA},
url = {https://doi.org/10.1145/3579371.3589350},
doi = {10.1145/3579371.3589350},
booktitle = {Proceedings of the 50th Annual International Symposium on Computer Architecture},
articleno = {82},
numpages = {14},
location = {Orlando, FL, USA},
series = {ISCA '23}
}

@online{Groq,
author = {Groq},
title = {GroqCard Accelerator},
url = {https://groq.com/wp-content/uploads/2024/02},
note = {2022}
}

@ARTICLE{EyerissV2,
  author={Chen, Yu-Hsin and Yang, Tien-Ju and Emer, Joel and Sze, Vivienne},
  journal={IEEE Journal on Emerging and Selected Topics in Circuits and Systems}, 
  title={Eyeriss v2: A Flexible Accelerator for Emerging Deep Neural Networks on Mobile Devices}, 
  year={2019},
  volume={9},
  number={2},
  pages={292-308},
  doi={10.1109/JETCAS.2019.2910232}}

@article{fang2024energy,
  title={An Energy-Efficient Unstructured Sparsity-Aware Deep SNN Accelerator With 3-D Computation Array},
  author={Fang, Chaoming and Shen, Ziyang and Wang, Zongsheng and Wang, Chuanqing and Zhao, Shiqi and Tian, Fengshi and Yang, Jie and Sawan, Mohamad},
  journal={IEEE Journal of Solid-State Circuits},
  year={2024},
  publisher={IEEE}
}

@article{thorpe1996speed,
  title={Speed of processing in the human visual system},
  author={Thorpe, Simon and Fize, Denis and Marlot, Catherine},
  journal={nature},
  volume={381},
  number={6582},
  pages={520--522},
  year={1996},
  publisher={Nature Publishing Group UK London}
}

@misc{prosperity,
      title={Prosperity: Accelerating Spiking Neural Networks via Product Sparsity}, 
      author={Chiyue Wei and Cong Guo and Feng Cheng and Shiyu Li and Hao "Frank" Yang and Hai "Helen" Li and Yiran Chen},
      year={2025},
      eprint={2503.03379},
      archivePrefix={arXiv},
      primaryClass={cs.AR},
      url={https://arxiv.org/abs/2503.03379}, 
}

@misc{phi,
      title={Phi: Leveraging Pattern-based Hierarchical Sparsity for High-Efficiency Spiking Neural Networks}, 
      author={Chiyue Wei and Bowen Duan and Cong Guo and Jingyang Zhang and Qingyue Song and Hai "Helen" Li and Yiran Chen},
      year={2025},
      eprint={2505.10909},
      archivePrefix={arXiv},
      primaryClass={cs.AR},
      url={https://arxiv.org/abs/2505.10909}, 
}

@misc{SEENN,
      title={SEENN: Towards Temporal Spiking Early-Exit Neural Networks}, 
      author={Yuhang Li and Tamar Geller and Youngeun Kim and Priyadarshini Panda},
      year={2023},
      eprint={2304.01230},
      archivePrefix={arXiv},
      primaryClass={cs.NE},
      url={https://arxiv.org/abs/2304.01230}, 
}

@article{furber2014spinnaker,
  title={The spinnaker project},
  author={Furber, Steve B and Galluppi, Francesco and Temple, Steve and Plana, Luis A},
  journal={Proceedings of the IEEE},
  volume={102},
  number={5},
  pages={652--665},
  year={2014},
  publisher={IEEE}
}

@article{OEDSNN,
   title={Optimizing event-driven spiking neural network with regularization and cutoff},
   volume={19},
   ISSN={1662-453X},
   url={http://dx.doi.org/10.3389/fnins.2025.1522788},
   DOI={10.3389/fnins.2025.1522788},
   journal={Frontiers in Neuroscience},
   publisher={Frontiers Media SA},
   author={Wu, Dengyu and Jin, Gaojie and Yu, Han and Yi, Xinping and Huang, Xiaowei},
   year={2025},
   month=feb }

@misc{SpikeCP,
      title={Knowing When to Stop: Delay-Adaptive Spiking Neural Network Classifiers with Reliability Guarantees}, 
      author={Jiechen Chen and Sangwoo Park and Osvaldo Simeone},
      year={2024},
      eprint={2305.11322},
      archivePrefix={arXiv},
      primaryClass={cs.NE},
      url={https://arxiv.org/abs/2305.11322}, 
}

@inproceedings{huang2011high,
  title={A high-performance, high-density 28nm eDRAM technology with high-K/metal-gate},
  author={Huang, KC and Ting, YW and Chang, CY and Tu, KC and Tzeng, KC and Chu, HC and Pai, CY and Katoch, A and Kuo, WH and Chen, KW and others},
  booktitle={2011 International Electron Devices Meeting},
  pages={24--7},
  year={2011},
  organization={IEEE}
}

@inproceedings{redmon2016you,
  title={You only look once: Unified, real-time object detection},
  author={Redmon, Joseph and Divvala, Santosh and Girshick, Ross and Farhadi, Ali},
  booktitle={Proceedings of the IEEE conference on computer vision and pattern recognition},
  pages={779--788},
  year={2016}
}

@article{jain2020rxnn,
  title={RxNN: A framework for evaluating deep neural networks on resistive crossbars},
  author={Jain, Shubham and Sengupta, Abhronil and Roy, Kaushik and Raghunathan, Anand},
  journal={IEEE Transactions on Computer-Aided Design of Integrated Circuits and Systems},
  volume={40},
  number={2},
  pages={326--338},
  year={2020},
  publisher={IEEE}
}

@inproceedings{lin2014microsoft,
  title={Microsoft coco: Common objects in context},
  author={Lin, Tsung-Yi and Maire, Michael and Belongie, Serge and Hays, James and Perona, Pietro and Ramanan, Deva and Doll{\'a}r, Piotr and Zitnick, C Lawrence},
  booktitle={European conference on computer vision},
  pages={740--755},
  year={2014},
  organization={Springer}
}

@article{everingham2010pascal,
  title={The pascal visual object classes (voc) challenge},
  author={Everingham, Mark and Van Gool, Luc and Williams, Christopher KI and Winn, John and Zisserman, Andrew},
  journal={International journal of computer vision},
  volume={88},
  number={2},
  pages={303--338},
  year={2010},
  publisher={Springer}
}

@article{fouda2020ir,
  title={IR-QNN framework: An IR drop-aware offline training of quantized crossbar arrays},
  author={Fouda, Mohammed E and Lee, Sugil and Lee, Jongeun and Kim, Gun Hwan and Kurdahi, Fadi and Eltawi, Ahmed M},
  journal={IEEE Access},
  volume={8},
  pages={228392--228408},
  year={2020},
  publisher={IEEE}
}

@article{lie2023cerebras,
  title={Cerebras architecture deep dive: First look inside the hardware/software co-design for deep learning},
  author={Lie, Sean},
  journal={Ieee Micro},
  volume={43},
  number={3},
  pages={18--30},
  year={2023},
  publisher={IEEE}
}

@article{spinnaker2,
  title={Spinnaker2: A large-scale neuromorphic system for event-based and asynchronous machine learning},
  author={Gonzalez, Hector A and Huang, Jiaxin and Kelber, Florian and Nazeer, Khaleelulla Khan and Langer, Tim and Liu, Chen and Lohrmann, Matthias and Rostami, Amirhossein and Sch{\"o}ne, Mark and Vogginger, Bernhard and others},
  journal={arXiv preprint arXiv:2401.04491},
  year={2024}
}

@inproceedings{hwang2026gustavsnn,
  title={GustavSNN: Unleashing the Power of Gustavson's Algorithm on SNN Acceleration with Column-Parallel Tick-Batch Dataflow},
  author={Hwang, Sangwoo and Lee, Donghun and Koo, Jahyun and Kung, Jaeha},
  booktitle={2026 IEEE International Symposium on High Performance Computer Architecture (HPCA)},
  pages={1--14},
  year={2026},
  organization={IEEE}
}

@misc{hala_point_intel2024,
  title        = {Intel Builds World’s Largest Neuromorphic System to Enable More Sustainable AI},
  author       = {{Intel Newsroom}},
  year         = {2024},
  howpublished = {\url{https://newsroom.intel.com/artificial-intelligence}},
  note         = {Accessed: 2026-04-26}
}

@inproceedings{liu202033,
  title={33.2 A fully integrated analog ReRAM based 78.4 TOPS/W compute-in-memory chip with fully parallel MAC computing},
  author={Liu, Qi and Gao, Bin and Yao, Peng and Wu, Dong and Chen, Junren and Pang, Yachuan and Zhang, Wenqiang and Liao, Yan and Xue, Cheng-Xin and Chen, Wei-Hao and others},
  booktitle={2020 IEEE International Solid-State Circuits Conference-(ISSCC)},
  pages={500--502},
  year={2020},
  organization={IEEE}
}

@inproceedings{polymorpic,
  author = {Zou, Cheng and Wei, Ziling and Lee, Jun Yan and Nie, Chen and You, Kang and He, Zhezhi},
title = {PolymorPIC: Embedding Polymorphic Processing-in-Cache in RISC-V based Processor for Full-stack Efficient AI Inference},
  booktitle = {2025 58th IEEE/ACM International Symposium on Microarchitecture (MICRO)},
  year = {2025},
doi={10.1145/3725843.3756066}
}

@ARTICLE{Nie2024VSPIM,
  author={Nie, Chen and Tang, Chenyu and Lin, Jie and Hu, Huan and Lv, Chenyang and Cao, Ting and Zhang, Weifeng and Jiang, Li and Liang, Xiaoyao and Qian, Weikang and Sun, Yanan and He, Zhezhi},
  journal={IEEE Transactions on Computers}, 
  title={VSPIM: SRAM Processing-in-Memory DNN Acceleration via Vector-Scalar Operations}, 
  year={2024},
  volume={73},
  number={10},
  pages={2378-2390},
  doi={10.1109/TC.2023.3285095}}

@inproceedings{MAICC,
author = {Fan, Renhao and Cui, Yikai and Chen, Qilin and Wang, Mingyu and Zhang, Youhui and Zheng, Weimin and Li, Zhaolin},
title = {MAICC: A Lightweight Many-core Architecture with In-Cache Computing for Multi-DNN Parallel Inference},
year = {2023},
isbn = {9798400703294},
publisher = {Association for Computing Machinery},
address = {New York, NY, USA},
url = {https://doi.org/10.1145/3613424.3614268},
doi = {10.1145/3613424.3614268},
booktitle = {Proceedings of the 56th Annual IEEE/ACM International Symposium on Microarchitecture},
pages = {411–423},
numpages = {13},
location = {Toronto, ON, Canada},
series = {MICRO '23}
}

\end{document}